\documentclass[a4paper]{ar-astro2e}

\usepackage{natbib}  

\usepackage{amssymb}
\usepackage{amsmath}
\usepackage{color}
\usepackage{url}
\usepackage{graphicx}

\def\aap{Astron. Astrophys.}

\def\aj{A. J.}
\def\apj{Ap. J.}

\def\apjs{Ap. J. Suppl.}

\def\nature{Nature}
\def\science{Science}
\def\pasj{Pub. Astron. Soc. Japan}

\def\aspc{Astron. Soc. Pac. Conf. Ser.}
\def\apss{Astrophys. Space Sci.}

\def\ie {i.e.,}
\def\eg {e.g.,}
\def\etal {et al.}

\def\uas {$\mu$as}
\def\masy{mas~y$^{-1}$}
\def\uasy{$\mu$as~y$^{-1}$}
\def\kms {km~s$^{-1}$}
\def\peryr {yr$^{-1}$}

\def\uv  {($u,v$)}
\def\xy  {($x,y$)}
\def\dxdy  {($\Delta x,\Delta y$)}
\def\clean {{\small CLEAN}}

\def\Msun {\ifmmode{M_\odot}\else{$M_\odot$}\fi}
\def\Lsun {\ifmmode{L_\odot}\else{$L_\odot$}\fi}

\def\HI  {H~{\small I}}
\def\hho {H$_2$O}
\def\sio {SiO}

\def\Ho  {\ifmmode{{\rm H}_0}\else{H$_0$}\fi}
\def\Ro  {\ifmmode{{\rm R}_0}\else{R$_0$}\fi}
\def\To  {\ifmmode{\Theta_0}\else{$\Theta_0$}\fi}

\def\lax{\mathrel{\rlap{\lower4pt\hbox{\hskip1pt$\sim$}}
    \raise1pt\hbox{$<$}}}                
\def\gax{\mathrel{\rlap{\lower4pt\hbox{\hskip1pt$\sim$}}
    \raise1pt\hbox{$>$}}}                

\def\tauLij     {\ifmmode {\tau(\nu_L)} \else {$\tau(\nu_L)$} \fi}
\def\tauHij     {\ifmmode {\tau(\nu_H)} \else {$\tau(\nu_H)$} \fi}

\def\tauDij     {\ifmmode {\tau^{ij}_{D}} \else {$\tau^{ij}_{D}$} \fi}
\def\tauijt     {\ifmmode {\tau^{ij}_{t}} \else {$\tau^{ij}_{t}$} \fi}
\def\tauit      {\ifmmode {\tau^{i}_{D_V}(t)} \else {$\tau^{i}_{D_V}(t)$} \fi}
\def\taujt      {\ifmmode {\tau^{j}_{D_V}(t)} \else {$\tau^{j}_{D_V}(t)$} \fi}
\def\tauitr     {\ifmmode {\tau^{i}_{D_V}(t_o)} \else {$\tau^{i}_{D_V}(t_o)$} \fi}
\def\taujtr     {\ifmmode {\tau^{j}_{D_V}(t_o)} \else {$\tau^{j}_{D_V}(t_o)$} \fi}

\def\deltauij   {\ifmmode {\Delta\tau^{ij}_{D}} \else {$\Delta\tau^{ij}_{D}$} \fi}

\begin{document}

\jname{Annu. Rev. Astronomy Astrophysics}
\jyear{2014}
\jvol{54}
\ARinfo{}

\title{Micro-Arcsecond Radio Astrometry}

\markboth{Reid \& Honma}{Radio Astrometry}

\author{M.J.~Reid
\affiliation{Harvard-Smithsonian Center for Astrophysics, 60 Garden Street, 
             \\Cambridge, MA 02138, U.S.A.}
        M.~Honma
\affiliation{Mizusawa VLBI Observatory, National Astronomical Observatory
             \\ of Japan \& Department of Astronomical Science,
             \\The Graduate University for Advanced Study, 
             \\Mitaka 181-8588, Japan}
}


\begin{abstract}
Astrometry provides the foundation for astrophysics.  Accurate positions are 
required for the association of sources detected at different times or
wavelengths, and distances are essential to estimate the size, luminosity,
mass, and ages of most objects.  Very Long Baseline Interferometry at
radio wavelengths, with diffraction-limited imaging at sub-milliarcsec
resolution, has long held the promise of micro-arcsecond astrometry.
However, only in the past decade has this been routinely achieved.
Currently, parallaxes for sources across the Milky Way are 
being measured with $\sim10$ \uas\ accuracy and proper motions of
galaxies are being determined with accuracies of $\sim1$ \uasy.  
The astrophysical applications of these measurements cover many fields, 
including star formation, evolved stars, stellar and super-massive black holes, 
Galactic structure, the history and fate of the Local Group, the Hubble 
constant, and tests of general relativity.  This review summarizes the 
methods used and the astrophysical applications of micro-arcsecond radio 
astrometry.
\end{abstract}

\maketitle

\section {Introduction}

\subsection {History}

For centuries astrometry was the primary focus of astronomy.  
Before the 17$^{th}$ century,
astronomers charted the locations of naked-eye stars with an accuracy 
of a fraction of an arcminute.  With this level of accuracy, all but 
a handful of the highest proper motion stars remain ``fixed'' for 
an astronomer's lifetime, and it was not until Galileo's introduction of
the telescope in astronomy that arcsecond precision became a possibility.

For millennia two great questions remained unanswered: was the
Earth at the center of the Universe and how large was the Universe?
Both questions could be answered through astrometry by measuring
the parallax of stars.  If the Earth revolved around the Sun, 
stars would exhibit yearly shifts in apparent position (annual parallax) and, 
if measured, these would indicate their distances (at that time equivalent 
to the ``size of the Universe'').  This scientific opportunity led
most scientists of the period to attempt astrometric measurements
to detect parallax (\citealt{Hirshfeld:01}).  

Early attempts to measure parallax by Robert Hooke, and later 
James Bradley, compared the tilt of a near-vertical telescope
to that of a plumb-line as the star $\gamma$~Draconis transited
directly overhead in London.   Bradley, after discovering and accounting
for aberration of light (a yearly $\pm$20~arcsec effect caused by the
Earth's orbital motion), was only able to place an upper limit of
0.5 arcsec on the annual parallax of the star.  It was not until 1838 
that Friedrich Wilhelm Bessel presented the first convincing measurement
of stellar parallax: 0.314 arcsec for the star 61~Cygni (chosen 
as a candidate by its large proper motion of $\approx5$ arcsec yr$^{-1}$).

Bessel and others followed a suggestion attributed to Galileo to measure 
{\it relative} parallax, the differential position shift of a nearby target 
star relative to a distant background star, rather than {\it absolute}
position shifts.  This approach still forms the basis of the highest accuracy 
measurements made today.  Currently astrometric accuracy using 
Very Long Baseline Interferometry (VLBI) at centimeter wavelengths is
approaching the $\sim1$ \uas\ level.  This review documents the
state of the art in radio astrometry, focusing both on the techniques
and the scientific results.

\subsection {Radio Astrometry}

Karl Jansky made the first astronomical observations at radio
wavelengths in the 1920s.  While searching for the source of interference 
in trans-atlantic telephone calls (mostly from lightning in the tropics),
he noted strong emission localized within a few degrees in the 
constellation of Sagittarius.  The peak of the signal drifted in time 
at a sidereal rate, indicating an origin outside the Solar System, 
and was ultimately traced to energetic (synchrotron emitting) electrons 
throughout the Milky Way, but concentrated toward the center.   

The development of radar during WWII, gave a boost to radio astronomy
in the 1940s and 1950s.   Astrometric precision with a single radio
telescope was generally limited to some fraction of the diffraction
limit, $\theta_d \approx \lambda/D$, where $\lambda$ is the observing 
wavelength and $D$ is the antenna diameter.  Even though radio antennas
were an order of magnitude larger than optical telescopes, the 
roughly four orders of magnitude longer wavelength seemed to doom
radio astrometry.   However, that changed with  
the development of radio interferometry.  In the 1960s, interferometers
with baselines of $\sim1$ km localized the positions of quasi-stellar
objects (QSOs), leading to the discovery of highly redshifted optical
emission.  Over the years, with increasing baseline length, positional
accuracy with (connected-element) interferometers improved from $\sim1$
to $\sim0.03$ arcseconds (\citealt{Wade:77}).

In the late 1960s, radio interferometry was greatly extended by 
removing the need for a direct connection (either with cables or
microwave links) between antennas.  Signals were accurately time 
tagged using independent atomic clocks and recorded on 
magnetic tape, allowing separations of interferometer elements
across the Earth.  This technique, called Very Long Baseline Interferometry
(VLBI), lead to many discoveries, including super-luminal motion in
jets from active galactic nuclei (AGN) and upper limits of $\sim1$ pc 
on the size of the emitting regions.  Both results provided strong
evidence for super-massive black holes as the engines for AGN (\citealt{ReidIJMPD}).
Early VLBI observations with intrinsic angular resolution better than 1 mas
offered absolute position accuracy of $\sim0.3$ mas using
group-delay observables (\citealt{Clark:76,Ma:86}) and relative 
astrometric accuracy between fortuitously close pairings of 
QSOs of $\sim10$ \uas\ using phase-delay information (\citealt{Marcaide:85}).   

Starting in the 2000s, calibration techniques improved to the point where 
relative positional accuracy of $\sim10$ \uas\ could be routinely 
achieved for most bright targets, relative to a detectable QSO usually
within a couple of degrees on the sky.   This changed the game dramatically.
Such astrometric accuracy is unsurpassed in astronomy and is comparable to, 
or better than, the target accuracy of the next European astrometric space 
mission: Gaia (\citealt{Bourda:11}). 
Since radio waves are not absorbed significantly by interstellar dust,
the entire Milky Way is available for observation.  Also, since QSOs are
used as the position reference, {\it absolute} parallax and proper motions
are directly measurable.  Such measurements for over 100 star forming
regions have now been made for sources as distant as 11 kpc.
Significant results (see \S\ref{sect:applications}) include a resolution of the
Hipparcos Pleiades distance controversy (\citealt{Melis:14}), 
3-dimensional ``imaging'' of nearby star forming regions (\citealt{Loinard:07}), 
and the most accurate measurements to date of the distance to the Galactic 
Center, \Ro, and the circular rotation speed of the Local Standard of Rest,
\To\ (\citealt{Reid:14}).

\section{Astrophysical Applications} \label{sect:applications}

Since distance is fundamental to astrophysical understanding, it should not
be surprising that radio astrometry and parallax measurements are critical
for characterizing a wide variety of phenomena and classes of sources.  
Here we briefly discuss some of the more important astrophysical applications 
of radio astrometry.

\subsection {Galactic Structure} \label{sect:galactic}

\begin{figure}[h]
  \center{\includegraphics[width=0.9\textwidth]
          {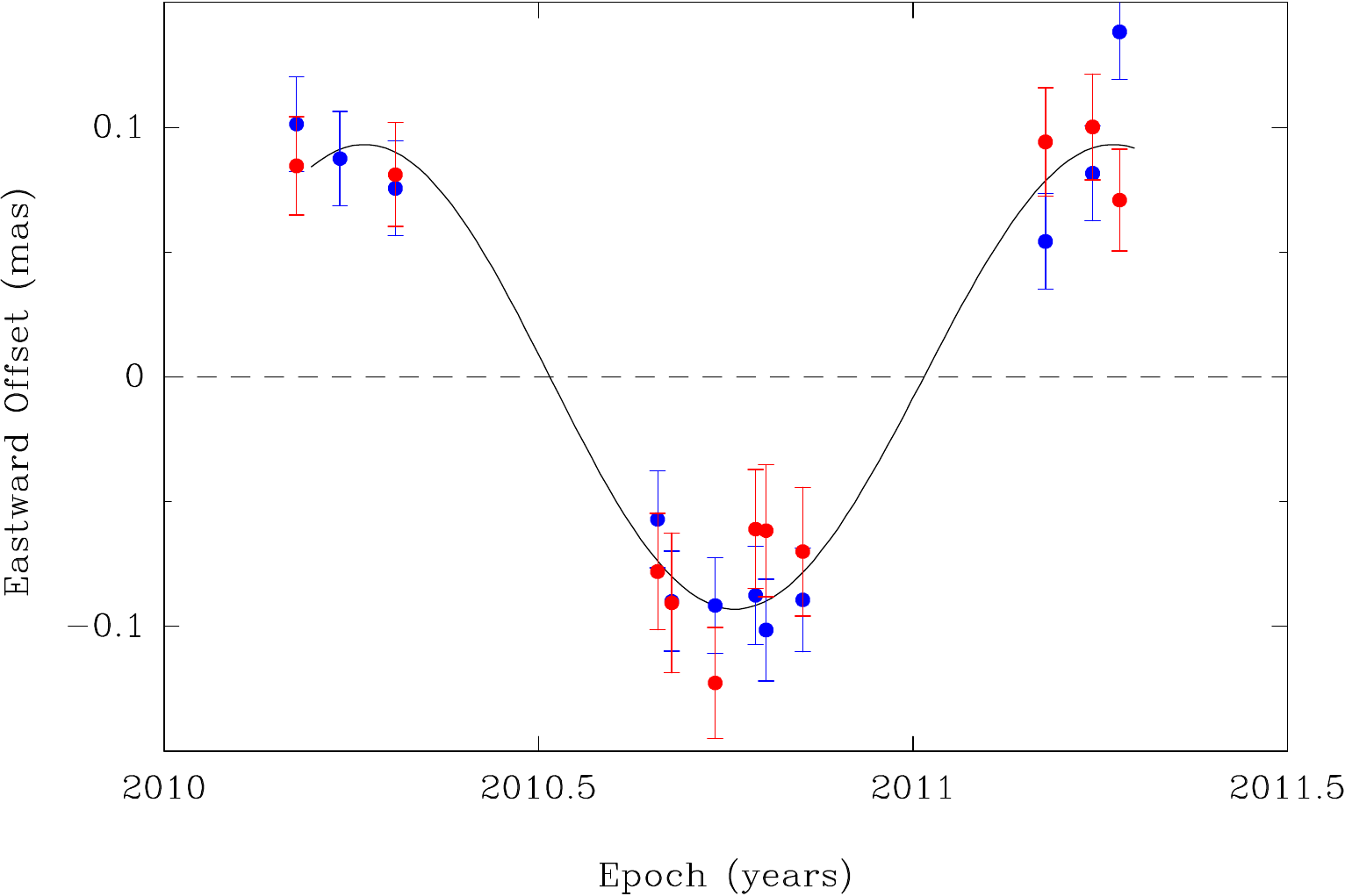}
         }
  \caption{Parallax data for W~49N measured with the VLBA after \citet{Zhang:13}.  
          Plotted are eastward position offsets versus time for \hho\ 
          maser spots at an LSR velocity of 8.3 \kms\ relative to the background
          quasar J1905+0952 (shown in red) and a maser spot at an 
           LSR velocity of 4.9 \kms\ relative to the background
          quasar J1922+0841 (shown in blue).
          The best fit proper motion has been removed, allowing the data         
          to be overlaid and effects of parallax to be more clearly seen.  
          This source has a parallax of $0.090\pm0.006$ mas,
          corresponding to a distance of $11.1\pm0.8$ kpc.
          }
 \label{fig:W49}
\end{figure}

Surprisingly, we know little of the spiral structure of the Milky Way;
there is considerable debate over the number of spiral arms, the nature of 
the central bar, and the values of the fundamental parameters \Ro\ (distance 
to Galactic center) and \To\ (circular rotation speed of the LSR).  Major
projects are underway to map the Milky Way with the VLBI Exploration of Radio 
Astrometry (VERA) array and the Very Long Baseline Array (VLBA) array,
though the Bar and Spiral Structure Legacy (BeSSeL) survey.  
With typical parallax accuracy of $\pm20$~\uas, and best accuracy of $\pm5$~\uas, 
one can measure distances of 5 and 20 kpc, respectively, with $\pm$10\% accuracy.
For example, parallax data for the most distant source measured to date, 
the massive star forming region W~49, are shown in {\bf Figure~\ref{fig:W49}}.
Over 100 distances to high-mass star forming regions have been measured 
using the astronomical ``gold standard'' technique of trigonometric parallax
(see \citet{Reid:09b,Honma:12,Reid:14} and references therein).

\begin{figure}[h]
  \center{\includegraphics[width=0.9\textwidth]
          {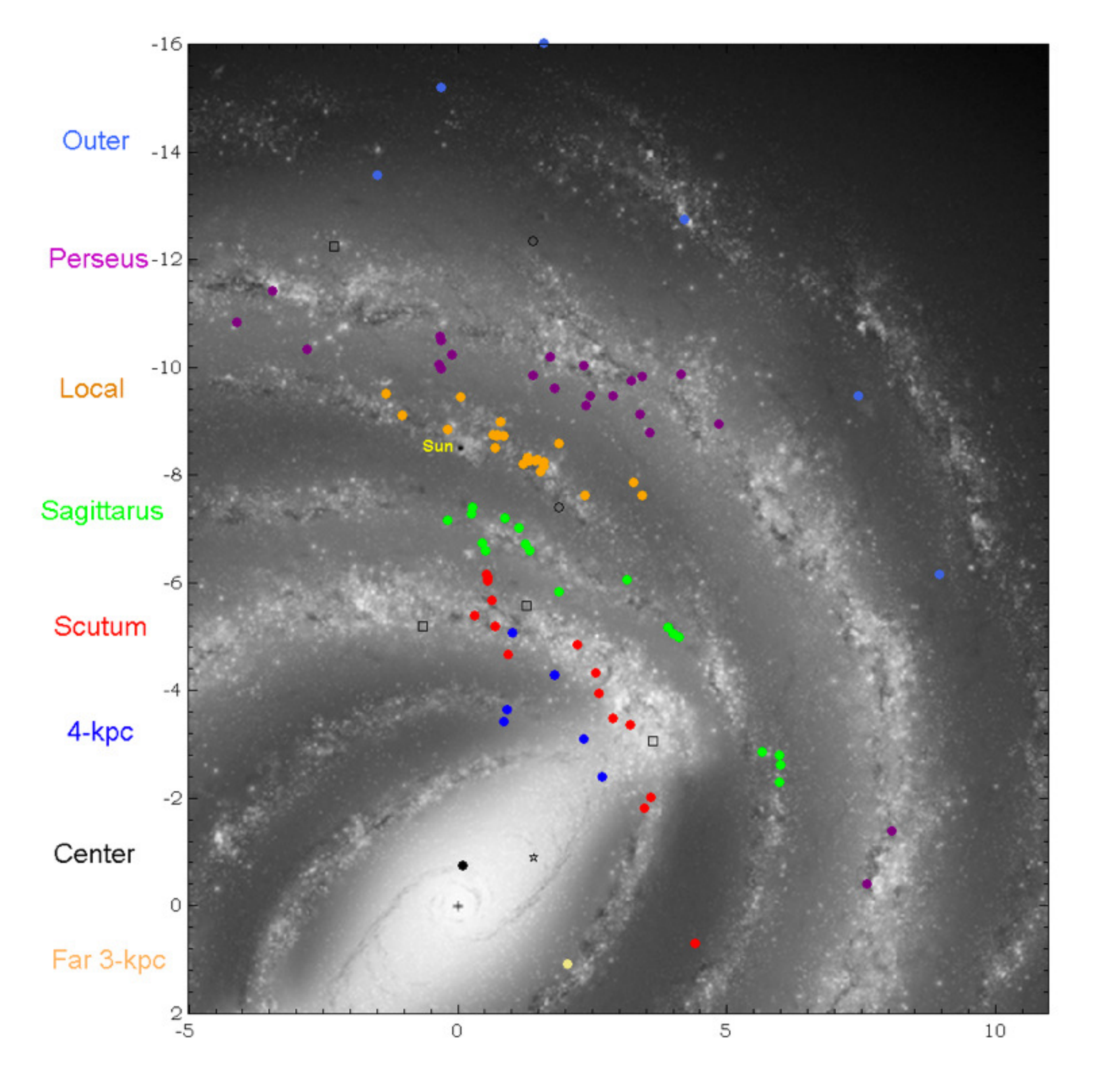}
         }
  \caption{Plan view of Milky Way.  The background is an artist conception, 
           guided by VLBI astrometry and Spitzer Space Telescope photometry (R. Hurt: IPAC).
           Dots are the locations of newly formed OB-type stars determined
           from trigonometric parallaxes using associated maser emission.
           The parallaxes were determined with the VLBA, VERA, and the EVN.
           Assignment to spiral arms (indicated by dot color) has been
           made by comparison to large scale emission of carbon monoxide
           in Galactic longitude--Doppler velocity space, independent of 
           distance.  The Galactic center is denoted by the plus (+) sign
           at (0,0) kpc; the Sun is labeled in yellow at (0,8.4) kpc.  
           On this view, the Milky Way rotates clockwise.  
          }
 \label{fig:parallaxes}
\end{figure}

Attempts to decode the nature of the spiral structure of the Milky Way have 
long relied on kinematic distance estimates (by comparing the Doppler velocity of 
a source and that expected from a model of Galactic rotation with 
distance as a free parameter).  One of the first high-precision VLBA parallaxes
was for the massive star forming region W3(OH); \citet{Xu:06} found a
distance of $1.95\pm0.04$ kpc, confirmed by \citet{Hachisuka:06}.  This distance
was more than a factor of two less than its kinematic distance, demonstrating
how unreliable kinematic distances can be.  Now, with ``gold standard'' trigonometric 
parallaxes, the major spiral features of the Milky Way are, for the first time, being 
accurately located and spiral arm pitch angles measured 
(see {\bf Figure~\ref{fig:parallaxes}}).  
Recent surprises include that the Local (Orion) arm, thought to be a minor structure,
is longer and has far more on-going star formation than previously thought
and rivals the Perseus spiral arm in the second and third Galactic quadrants
(\citealt{Xu:13}).

With measured source coordinates, distance, proper motion, and radial velocity,
one has full phase-space (3-dimensional position and velocity) information.
These data can be modeled to yield an estimate of \Ro, \To, and the slope of the
Galaxy's rotation curve.  The latest modeling indicates that $\Ro=8.35\pm0.16$~kpc and
$\To=251\pm8$~\kms\ (\citealt{Reid:14}).  Also, the rotation curve between 4 and 13 kpc
from the Galactic center is very flat (\citealt{Honma:07,Reid:14}).  
Such a large value for \To\ (about
15\% larger than the IAU recommended value of 220~\kms), if confirmed by other 
measurements, would have widespread impact in astrophysics, including
increasing the total mass (including dark matter halo) of the Milky Way by
about 50\%, revising Local Group dynamics by changing the Milky Way's mass 
and velocities of Group members (when transforming from Heliocentric to
Galactocentric coordinate systems), and increasing the expected signal from 
dark matter annihilation radiation.  

\subsection {Star Formation}

Gould's Belt, a flattened structure of star forming regions of radius
$\approx~1$ kpc and centered $\approx0.1$ kpc from the Sun (toward the Galactic 
anticenter), contains most of the sites of current star formation 
near the Sun (\eg\ the Ophiucus, Lupus, Taurus, Orion, Aquila Rift and
Serpens star forming regions).  Most of our knowledge about the
formation of stars like the Sun comes from in-depth studies of these 
regions, for example, from the Spitzer c2d survey (\citealt{Evans:09}),
the XMM Newton Extended Survey of Taurus (\citealt{Guedel:08}) and
the Herschel Gould Belt Survey (\citealt{Andre:10}).  
Of course, knowledge of distance is critical for quantitative measures 
of cloud and young stellar object (YSO) sizes, masses, luminosities and ages.  
For such deeply embedded sources, which are optically invisible, typically 
distances had not been estimated to an accuracy of better than $\pm30$\%.  
Since some physical parameters depend on distance squared or cubed, these 
parameters can be in error by factors of $\approx2$.   

Trigonometic parallax measurements with VLBI techniques can be made
by observing gyro-synchrotron emission from T Tauri objects, which
is usually confined to a region of a few stellar radii, or from
\hho\ maser emission often associated with Herbig-Haro outflows.  
Currently, about a dozen YSOs parallaxes have been obtained, with 
up to 200 planned in this decade.

In the Ophiucus cloud, \citet{Loinard:08}, using the VLBA, found that 
sources S1 and DoAr21 are at a mean distance of $120.0\pm4.5$ pc; 
\citet{Imai:07} using VERA found a consistent distance of $178^{+18}_{-37}$ 
pc, albeit with larger uncertainty, for the \hho\ masers toward IRAS 16293$-$2422.
The Taurus molecular cloud has been well studied and parallaxes for
five YSOs have been reported from VLBA observations in a series of papers
(\citealt{Loinard:05,Loinard:07,Torres:07,Torres:09,Torres:12}).  These
locate three YSOs associated with the L~1495 dark cloud at a
distance of $131.4\pm1.4$ pc and, interestingly, T Tauri Sb at 
$146.7\pm0.6$ pc and HP Tau/G3 at $161.9\pm0.9$ pc.  
Clearly, these observations are tracing the 3-dimensional structure of 
the cloud, which is extended over about 30 pc along the line of sight, 
consistent with its extent on the sky of $10^\circ$ ($\approx25$ pc).

VERA observations have yielded parallaxes to YSOs in the Perseus molecular 
cloud.  Observing \hho\ masers, \citet{Hirota:08} measured 
parallax distances of $235\pm18$ pc for the SVS~13 toward 
NGC~1333 and $232\pm18$ pc for L~1448~C (\citealt{Hirota:11}).
These measurements clarified the distance to the Perseus cloud, which
was highly uncertain -- estimated at between 220 pc (\citealt{Cernis:90})
and 350 pc (\citealt{Herbig:83}).
The Serpens cloud provides another example where previous optical-based
distance estimates varied considerably from 250 to 700 pc, with recent
convergence toward the low end at $\sim230\pm20$ pc, 
as summarized by \citet{Eiroa:08}.
However, \citet{Dzib:10} used the VLBA to obtain a trigonometric parallax
distance for EC~59 of $414.9\pm4.4$ pc and suggested that faulty identification
of dusty clouds in the foreground Aquila Rift might account for optical 
distance estimate. 

The Orion Nebula is perhaps the most widely studied region of star formation.
Prior to 1981, distance estimates for the Orion Nebula ranged from about
380 to 520 pc as summarized by \citet{Genzel:81}.   By comparing 
radial and proper motions (measured with VLBI observations) of \hho\ masers 
toward the Kleinman--Low (KL) Nebula, an active region of star formation within 
the Orion Nebula, Genzel \etal\ estimated the distance to be $480\pm80$ pc.
Since then, four independent trigonometric parallax measurements have been performed.
\citet{Sandstrom:07} observed gyro-synchrotron emission from a YSO over
nearly 2 yr and obtained a parallax distance of $389\pm24$ pc using the VLBA.
\citet{Hirota:07} using VERA measured \hho\ masers and determined a
distance of $437\pm19$ pc for the KL region.
Recently, two very accurate parallaxes have been published:
\citet{Menten:07} used the VLBA and observed continuum emission from three YSOs and
determined a distance of $414\pm7$ pc; \citet{Kim:08} using VERA and observing
\sio\ masers estimated a distance of $418\pm6$ pc.  The latter two measurements
from different groups, using different VLBI arrays, different target sources, 
and different correlators and software are in excellent agreement.  
Together they indicate that the KL nebula/Trapezium region of the Orion Nebula is at 
a  distance of $416\pm5$pc, nearly a 1\% accurate distance!
Since, the Orion Nebula is the subject of large surveys, from x-rays to radio waves,
having such a ``gold-standard'' distance will enable precise estimates
of sizes, luminosities, masses, and ages.

A large number ($>100$) of parallaxes have been measured with the VLBA and EVN 
arrays for maser sources in high mass star forming regions 
(\citealt{Bartkiewicz:08,Brunthaler:09,Hachisuka:09,Moellenbrock:09,Sanna:09,Xu:09,
Zhang:09,Rygl:10,Sato:10,Moscadelli:11,Xu:11,Sanna:12,Immer:13})
and by the VERA array
(\citealt{Sato08,Sato10,Oh10,Ando11,Honma11,Kurayama11,Matsumoto11,Motogi11,
Nagayama11a,Nagayama11b,Niinuma11,Shiozaki11,Sakai12}).
As the focus of these observations has been to better understand Galactic structure, 
we discuss these in \S \ref{sect:galactic}.

\subsection{Asymptotic Giant Branch Stars}

Asymptotic Giant Branch (AGB) stars are in late stages of stellar
evolution and have large convective envelopes and high mass-loss rates.
They often exhibit maser emission from circumstellar OH, H$_2$O and SiO molecules,
providing good targets for radio astrometry with VLBI.
Accurate distances to AGB stars are necessary to constrain their
physical parameters, such as size and luminosity, and these are crucial
to test theories of late stages of stellar evolution.  And, of course,
distances are needed to calibrate the period-luminosity (P-L)
relation of Mira variables, which can be used as standard candles, 
\eg\ \citet{Whitelock08}.  Giant stars are not good astrometric targets
at optical wavelengths and the best optical parallaxes have accuracies 
poorer than a few mas.  However, parallaxes based on radio observations
of circumstellar masers has demonstrated more than a factor of ten
better parallax accuracy, allowing much better calibration of the 
Mira P-L relation.

Early astrometric observations with the VLBA for red-giant OH masers demonstrated
the potential for parallax measurements, achieving between 0.3 and 2 mas uncertainties
(\citep{Langevelde00,Vlemmings03}).  For example, parallax distances for S CrB 
($418^{+21}_{-18}$ pc) and U Her ($266^{+32}_{-28}$ pc) by \citet{Vlemmings07}
are considerably more accurate than those measured by the Hipparcos
satellite.   Astrometry of OH masers at the relatively low observing
frequency of 1.6 GHz are limited by uncompensated ionospheric effects,
but these effects can be minimized by observations near solar minimum and using
in-beam calibrators very close to the targets.

The first VLBI parallax of an AGB star using H$_2$O masers at 22 GHz
revealed a distance for UX Cyg, a long period Mira variable, 
of $1.85^{+0.25}_{-0.19}$ kpc (\citealt{kurayama05}).  Since then
the VERA array has been used to obtain parallaxes for many Mira and 
semi-regular variables, including S Crt (\citealt{Nakagawa08}), 
SY Scl \citep{Nyu11}, and RX Boo (\citealt{Kamezaki12}).  Of particular
interest are the astrometric measurements for the symbiotic system R Aqr 
(\citealt{Kamohara10}) and the parallax distance of $218^{+12}_{-11}$ pc
(\citealt{Min13}).  Future observations may enable one to trace
binary's orbital motion.

Red supergiants are rare objects, typically at kpc distances.
At these distances, and owing to their very large sizes and irregular
photospheres, they are beyond the reach of optical parallax measurements.
\citet{Choi08} observed \hho\ masers toward VY CMa, one of the
best-studied red supergiants, and found a parallax distance of 
$1.14^{+0.11}_{-0.09}$ kpc with VERA.  This distance was later
confirmed by \citet{Zhang12b}, who used the VLBA and found a 
distance of $1.20^{+0.13}_{-0.10}$ kpc.  Generally, the distance 
had been assumed to be 1.5 kpc, requiring an extraordinary
luminosity of $5\times10^5$ \Lsun; the parallax distances
reduce the luminosity estimate by 40\% to a more reasonable value. 
VLBI parallaxes accurate to $\pm10$\% have been obtained for several other 
red supergiants, including S Per (\citealt{Asaki10}), NML Cyg (\citealt{Zhang12b}), and 
IRAS 22480+6002 (\citealt{Imai12}).

Proto-Planetary Nebulae are in the final stage of evolution of
intermediate-mass stars, linking the AGB and planetary nebula phases. 
These sources are known to exhibit bipolar outflows with a velocities 
exceeding $\sim$100 km s$^{-1}$; they can have strong \hho\ masers
and are often called ``water-fountains.''  Parallaxes have been measured
for  IRAS 19134+2131 (\citealt{Imai07}, IRAS 19312+1950 (\citealt{Imai11}), 
IRAS 18286-0959 (\citealt{Imai13}), and K~3-35 (\citealt{Tafoya11}).

\subsection {X-ray Binaries}

The first trigonometric parallax for a black hole candidate was for
the X-ray binary V404 Cyg (\citealt{MillerJones:09}), indicating a
distance of $2.39\pm0.14$ kpc.  This value was significantly lower
than previously estimated and indicated that its 1989 outburst 
was not super-Eddington.  Its peculiar velocity, derived from the
radio proper motion, is only about 40 \kms, suggesting that it
did not receive a large natal ``kick'' from an asymmetric supernova
explosion.  This differs from many pulsars, which often have  
order of magnitude larger peculiar motions (see \S \ref{sect:pulsars}).

The long-standing uncertainty over the distance to Cyg X-1 (see, \eg, 
\citet{CaballeroNieves:09}) limited understanding of this famous
binary, including whether or not the unseen companion was a black hole.
Recently a trigonometric parallax measurement with the VLBA yielded a 
distance of $1.86\pm0.12$ kpc (\citealt{Reid:11}). 
Knowing the distance to the binary removed the mass--distance degeneracy
that limited the modeling of optical/IR data (light and velocity curves) and
revealed that the unseen companion in Cyg X-1 has a mass of $14.8\pm1.0$ \Msun\ 
(\citealt{Orosz:11}). This mass confidently exceeds the limit for a neutron star 
and firmly established it as a black hole.  Once the masses of the two
stars were accurately determined, X-ray data could be well modeled.  
This revealed that black hole spin is near maximal and, since the binary is too 
young for accretion to have appreciably spun up the black hole, most of the
spin angular momentum is probably natal (\citealt{Gou:11}).  Finally,
the 3-dimensional space-motion of the binary, from the radio astrometry and 
optical Doppler shifts, confirms Cyg X-1 as a member of the Cyg OB3 association,
and the lack of evidence for a supernova explosion in this region suggests
that the black hole may have formed via prompt collapse without an explosion
(\citealt{Mirabel:03}).

Recently, using the EVN and the VLBA, \citet{MillerJones:13} obtained a parallax 
distance of $114\pm2$ pc for SS Cyg, a binary composed of a white and red dwarf.   
This distance is significantly less than an optical parallax measured with the 
Hubble Space Telescope of $159\pm12$ pc (\citealt{Harrison:99}).  
The larger optical distance required the source to be significantly more 
luminous and proved difficult to reconcile with accretion disk theory.  
However, the smaller distance from radio astrometry seems to resolve this problem.

\subsection {Pulsars} \label{sect:pulsars}

Pulsar radio emissions are extremely compact and relatively bright,
which make them suitable for VLBI observations, and radio astrometry of
pulsars dates back to the 1980's.  The first interferometric parallax measurements 
were by \citet{Gwinn86}, who reported parallaxes for two pulsars
using a VLBI array which included the Arecibo 305-m telescope.
Later  \citet{Bailes90a} measured a parallax for PSR 1451$-$68 of $2.2\pm0.3$ mas, 
using the Parkes--Tidbinbilla Interferometer in Australia, and determined
a line-of-sight average interstellar electron density of $0.019\pm0.003$ cm$^{-3}$
by combining the distance and dispersion measures, suggesting that the
interstellar medium in the Solar neighborhood is typical of that over 
larger scales in the Galaxy.  Pulsar proper motions for 6 pulsars conducted with the 
Parkes--Tidbinbilla Interferometer revealed that motions away from the Galactic
plane were between 70 and 600 \kms\ (\citealt{Bailes90b}).
These early studies provided distances and  
confirmed the expectation that pulsars are high-velocity objects,
most-likely due to ``kicks'' associated with their parent supernova explosions.

Most recent pulsar astrometry uses ``in-beam'' calibrators to greatly improve
positional accuracy.  Other improvements include gating the pulsar signal 
during correlation to improve signal-to-noise ratios and performing ionospheric 
corrections based on multi-frequency phase fitting.
With these improvements and using the VLBA a parallax with better than 
$\pm10$\% accuracy was obtained for PSR B0950+08 at a distance of 280 pc 
(\citealt{Brisken00}), and nine other pulsar parallaxes with distances between
160 and 1400 pc have been reported (\citealt{Brisken02}). 

By observing at higher frequencies (5 GHz instead of 1.6 GHz), in order
to reduce the effects of the ionosphere, higher accuracy pulsar parallaxes 
have been obtained, for example, for PSR B0355+54 ($0.91\pm0.16$ mas) and 
PSR B1929+10 ($2.77\pm0.07$ mas) (\citealt{Chatterjee04}).
More recently, \citet{Chatterjee09} obtained results for 14
pulsars with the VLBA, including a parallax for the most distance pulsar
yet measured: PSR B1514+09 with a parallax of $0.13\pm0.02$ mas, corresponding 
to a distance of $7.2^{+1.3}_{-1.2}$ kpc.   With this sample, it is
clear that most pulsars are moving away from the Galactic plane with 
speeds of hundreds of \kms.

Astrometry also provides a unique opportunity to
constrain pulsar birth places through velocity and distance measurements.
For instance, \citet{Campbell96} used a global VLBI array to measure the 
parallax and proper motion for PSR B2021+51, which
ruled out the supernova remnant HB 21 as the origin of the pulsar,
as it implied an improbable low age of only 700 years.  Similar studies
have been performed by others (\citealt{Dodson03,Ng07,Chatterjee09,Bietenholz13}). 

Astrometric observations also provide information on the physical properties 
of pulsars.  For example, \citet{Brisken03} combined a VLBA parallax for
PSR B0656+14 with a thermal X-ray emission model to constrain the stellar 
radius between 13 and 20 km.
\citet{Deller12b} measured a parallax for the transitional millisecond pulsar 
J1023+0038 with the VLBA.  Their distance of
$1368^{+42}_{-39}$ pc was twice that predicted by the standard interstellar 
plasma model.  When combined with timing and optical observations 
of this binary system, the new distance indicated a mass of
$M\sim 1.71\pm0.1$ \Msun, suggesting that it is a recycled pulsar.
\citet{Deller09b} conducted astrometric observations of 
seven pulsars in the southern hemisphere using Australian Long Baseline Array (LBA).
Their new distance to PSR J0630$-$2834 required the efficiency of conversion of
spin-down energy to X-rays to be less than 1\%, an order of magnitude
lower than previous estimates using a less accurate distance. 

Magnetars, pulsars with extremely strong magnetic field, are
also interesting astrometric targets. 
\citet{Helfand07} observed XTE J1810$-$197 with the VLBA and
obtained a proper motion of 212 \kms, suggesting that this magnetar has a 
lower space motion than theoretical predicted (\citealt{Duncan92}).
Recently, \citet{Deller12a} found a similarly low space motion for the magnetar 
J1550$-$5418.

Finally, pulsar astrometry can be critical for testing the constancy of
physical parameters. Using the VLBA, \citet{Deller08} observed J0437$-$4715, a
milli-second binary pulsar with a white dwarf companion.
They obtained a very precise parallax distance of $156.3\pm1.3$ pc.
Comparing this to a kinematic distance from pulsar timing gives
strong limits on unmodeled accelerations, which provide a limit
on the constancy of the Gravitational constant of
$\dot{G}/G = (-5\pm26)\times10^{-13}$ \peryr.
This constraint is consistent with those obtained from lunar laser
ranging as well as gravitational wave backgrounds.

\subsection {Radio Stars}

The Algol system is an eclipsing binary (with a
period of 2.9 days, consisting of B8 V primary and K0 IV secondary) and a
distant companion with an orbital period of 1.86 yr.
VLBI astrometry by \citet{Lestrade93} revealed that the radio emission
originates from the K0 subgiant and traced the orbital motion of eclipsing
binary, indicating that the orbit of the binary is
nearly orthogonal to that of the tertiary companion.
Observations of another hierarchical triple (Algol-like) system, UX Ari, by
\citet{Peterson11} detected the acceleration of the tight binary caused by the 
tertiary star. This acceleration measurement dynamically constrains the mass of 
the tertiary to be $\approx0.75$ \Msun, a value consistent with a spectroscopic
identification of a K1 main-sequence star.

Astrometric observations of radio stars allow one to accurately tie the 
fundamental radio and the optical reference frames.
A comparison between the radio and preliminary Hipparcos frames
by \citet{Lestrade95} revealed systematic discrepancies that could be
removed by a global rotation (and its time derivative).
Further VLBI observations by \citet{Lestrade99} achieved parallax accuracies 
of $\approx0.25$ mas for nine sources, whose distances ranged from 
20 to 150 pc.

\citet{Dzib13} monitored radio emission from colliding winds in Cyg OB2\#5
and obtained a marginal parallax of 0.61$\pm$0.22 mas, 
consistent with other distance measurements of Cyg X regions
(\citealt{Rygl12,Zhang12b}).  These observations also revealed a
high radio-brightness temperature ($\gax10^7$ K),
providing information for modeling the stellar winds.

\subsection {Star Clusters}

The Hyades and Pleiades clusters play a pivotal role in quantitative astrophysics,
serving as pillars of the astronomical distance ladder.  Recently, the
Hipparcos space mission, which measured $\approx100,000$ stellar parallaxes with 
typical accuracies of $\pm1$ mas, presented a (revised) parallax distance of 
$120.2\pm1.9$ pc for the Pleiades (\citealt{vanLeeuwen:09}).  This result has been quite
controversial, since a variety of other techniques, including main-sequence fitting,
generally give distances between 131 and 135 pc.  Using the HST fine guidance sensor, 
\citet{Soderblom:05} measured {\it relative} trigonometric parallaxes for three 
Pleiads, which, after correction to absolute parallaxes, average to a distance of 
$134.6\pm3.1$ pc.

In an attempt to provide a totally independent and straight-forward distance to
the Pleiades, \citet{Melis:14}  have used the VLBA with the Green Bank and Arecibo
telescopes to measure {\it absolute} parallaxes to several Pleiads that display compact 
radio emission.  Preliminary results suggest a cluster parallax 
near 138 pc, with an uncertainty of less than $\pm2$ pc (including
measurement error and cluster depth effects).  This result seems
to rule out the Hipparcos value, and it may even be in some tension with the
ensemble of astrophysical-based distance indicators.  Since the source of
error for the Hipparcos parallax for the Pleiades has not been convincingly established,
there could be concern for the Gaia mission, which is targeting a parallax accuracy of 
$\pm20$ \uas, since Gaia might inherit some unknown systematics from Hipparcos.  
Intercomparison of high accuracy VLBA parallaxes with those from Gaia will
provide a critical cross checking.

\subsection {Sgr A*}

Sgr A*, the candidate super-massive black hole (SMBH) at the center of the
Galaxy, is a strong radio source.  It is precluded from optical view by $>20$ mag
of visual extinction but can sometimes be detected when flaring at 2.2 $\mu$m wavelength
(through $\lax3$ mag of extinction at this wavelength).  Astrometric observations
in the infrared of stars orbiting an unseen mass have provided compelling
evidence of a huge mass concentration, almost surely a black hole 
(\citealt{Ghez:08,Gillessen:09}).  These observations require a grid of sources with 
accurate positions relative to Sgr A* to calibrate the infrared plate scale, rotation, 
and low-order distortion terms.  The calibration sources have been provided by 
VLA and VLBA astrometric observations relative to the radio bright Sgr A* 
of \sio\ and \hho\ masers in the circumstellar envelopes of red giant and supergiant 
stars within the central pc (\citealt{Reid:03}).  
This allowed location of the position of Sgr A* on infrared images, which matched that 
of the gravitational focal position of the stellar orbits to $\approx1$ mas accuracy, 
as well as confirming Sgr A*'s extremely low luminosity (\citealt{Menten:97}).

Astrometric observations of Sgr A* at 7~mm wavelength, relative to background 
quasars with the VLBA, yielded the angular motion of the Sun about the Galactic
center and placed extremely stringent limits on the mass density of the SMBH candidate.
While stars orbiting Sgr A* have been observed to move with speeds exceeding 
5000 \kms\ (\citealt{Schoedel:02}), the velocity component perpendicular to the 
Galactic plane of Sgr A* is less than $\approx1$ \kms, requiring that most of the
$4\times10^6$~\Msun\ required by the stellar orbits is tied to the radiative
source Sgr A* (\citealt{Reid:04}).   Since the millimeter wavelength emission 
from Sgr A* is confined to a $\sim1$ Schwarzschild radii region (\citealt{Doeleman:08}), 
the implied mass density is approaching that theoretically expected for a black hole
(\citealt{ReidIJMPD}).

\subsection {Megamasers and the Hubble Constant}

The Hubble constant, \Ho, is a critical cosmological parameter, not only for
the extragalactic distance scale, but also for determining the flatness of the
Universe and the nature of dark energy.  Some active galactic nuclei (AGN) with thin,
edge-on accretion disks surrounding their central super-massive black hole (SMBH)
exhibit \hho\ ``megamaser'' emission, with bright maser spots coming from clouds
in Keplerian orbit about the SMBH.  Astrometric observations can be used to
map the positions and velocities of these clouds.  Coupled with time monitoring
of the maser spectra, which allows direct measurement of cloud accelerations
(by tracking the velocity drift over time of maser features), one can obtain a geometric
(angular-diameter) distance estimate for the galaxy.  

Observations of the archetypal megamaser galaxy, NGC~4258, demonstrated
the power of radio astrometric observations for better understanding
of AGN accretion disks (\citealt{Herrnstein:05}) and yielded an accurate distance 
of $7.2\pm0.5$~Mpc to the galaxy (\citealt{Herrnstein:99}).  Since NGC~4258
is nearby and its (unknown) peculiar motion is likely to be a large
fraction of its cosmological recessional velocity ($\approx500$ \kms), one
cannot directly measure \Ho\ by dividing velocity by distance.  However, since
Cepheid variables have been observed in NGC~4258, one can
use the radio distance to recalibrate the zero-point of the
Cepheid period-luminosity relation (\citealt{Macri:06}) and then revise
estimates of \Ho\ (\citealt{Riess:12a,Riess:12b}).  Recently, \citet{Humphreys:13}
analyzed a decade of observations of NGC~4258 and, via more detailed modeling of
disk kinematics, obtained an extremely accurate distance of $7.60\pm0.23$~Mpc,
which constrains $\Ho=72.0\pm3.0$ \kms\ Mpc$^{-1}$. 

By observing \hho\ masers in galaxies like NGC~4258, but more distant and
well into the ``Hubble flow,'' the Megamaser Cosmology Project (MCP) is obtaining 
direct estimates of \Ho.  For galaxies in the Hubble flow, unknown
galaxy peculiar motions are a small source of systematic uncertainty ($\lax5$\%).
Results for the megamaser galaxy UGC~3789 (\citealt{Reid:09c,Braatz:10}) have yielded 
$\Ho=68.9\pm7.1$ \kms\ Mpc$^{-1}$ (\citealt{Reid:13}) and for NGC~6264  
$\Ho=68\pm9$ \kms\ Mpc$^{-1}$ (\citealt{Kuo:13}).  Based on these results, and
preliminary results for Mrk~1419, \citet{Braatz:13} find a combined result of
$\Ho=68.0\pm4.8$ \kms\ Mpc$^{-1}$.  The goal
of the MCP is measurement of $\approx10$ megamaser galaxies, each with
an accuracy near $\pm10$\%, which should yield a combined estimate of \Ho\ 
with $\pm3$\% accuracy.  While, formally, a $\pm3$\% uncertainty may be slightly
larger than claimed by other techniques, the megamaser method is direct 
(not dependent on standard candles) and totally independent. 

\subsection {Extragalactic Proper Motions}

In the concordance $\Lambda$CDM cosmological 
model, galaxies grow hierarchically by accreting smaller galaxies.  
Nearby examples of galaxy interactions are found in the environments of 
the Milky Way and the Andromeda Galaxy, the dominant galaxies in the Local Group.  
In the past, only {\it radial} velocities for Local Group galaxies were known and 
statistical approaches had to be used to model the system.  While the radial velocity 
of Andromeda indicates that it is approaching the Milky Way, without 
knowledge of its proper motion one cannot know, for example, if the two 
galaxies are on a collision course or if they are in a relatively stable 
orbit.

\begin{figure}[h]
  \center{\includegraphics[width=0.7\textwidth]
          {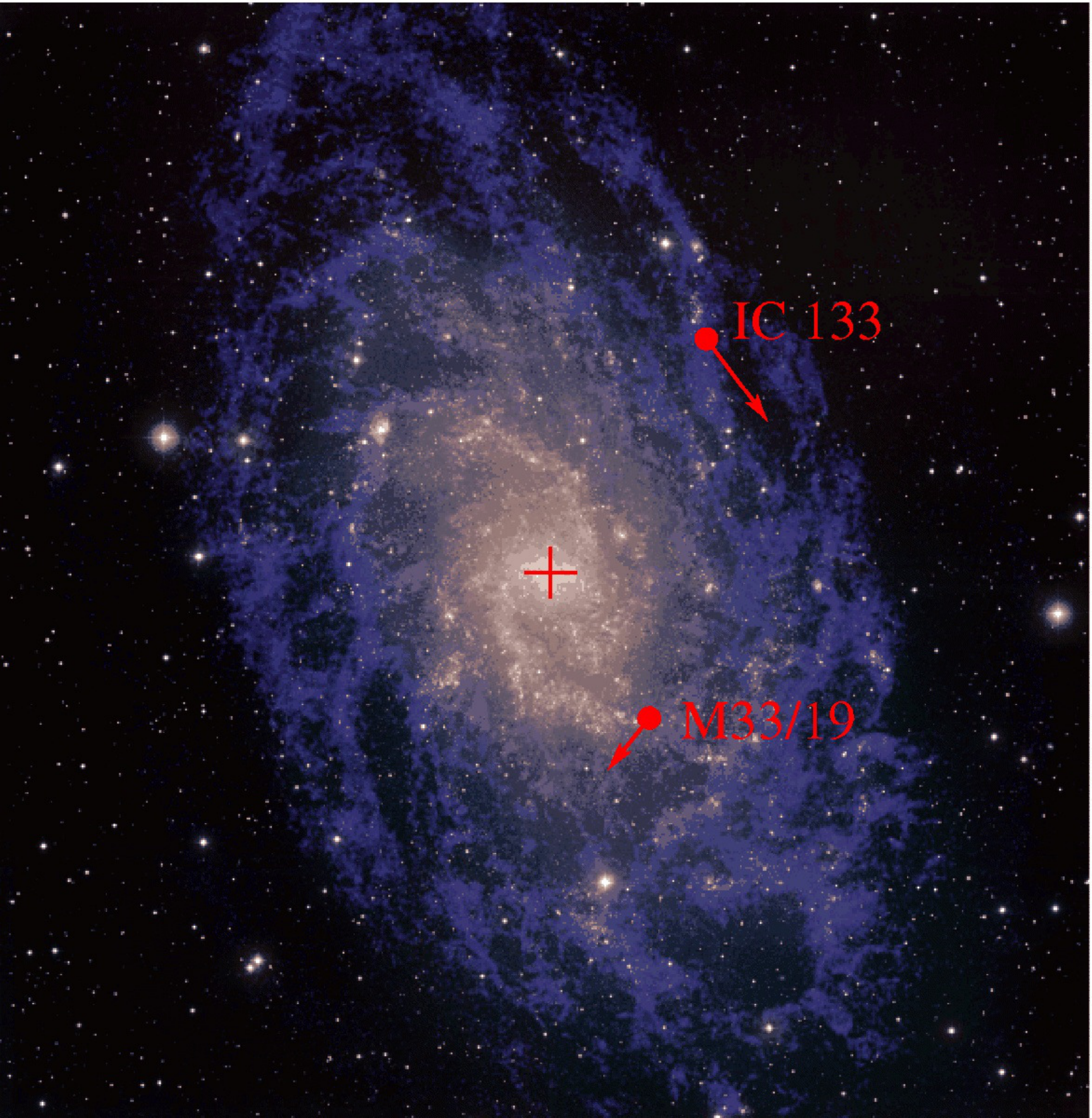}
         }
  \caption{Image of M 33, a satellite of the Andromeda galaxy, 
           with the locations and measured proper motions
           of \hho\ masers in two regions of massive star formation 
           (\citealt{Brunthaler:05}).  Both the relative motions
           (\ie\ the van Maanen experiment) and the absolute motions
           with respect to a background quasar have been measured.
           The relative motions gives a ``rotational parallax'' distance,
           and the absolute motion can be used to constrain the mass
           of the Andromeda galaxy.
          }
 \label{fig:M33}
\end{figure}

Astrometric VLBI observations have yielded both the internal angular rotation and 
the absolute proper motion of M~33, a satellite of Andromeda (\citealt{Brunthaler:05}).
The angular rotation of M33 was the focus of the van Maanen--Hubble
debate in the 1920's (\citealt{vanMaanen:23,Hubble:26}), with van Maanen
claiming an angular rotation of $20\pm1$ \masy.  Such a large angular
rotation required it to be nearby and part of the Milky Way, in order to
avoid implausibly large rotation speeds.  
Indeed, Shapley forwarded this as evidence that spiral nebulae were Galactic objects
in the famous Shapley--Curtis debate.  The angular rotation measured
by Brunthaler \etal\ is $\sim1000$ times smaller than van Maanen claimed,
and, of course, consistent with an external galaxy (see {\bf Figure~\ref{fig:M33}}).
Coupled with knowledge of the \HI\ rotation speed and inclination of M33,
the angular rotation rate yields a direct estimate of distance (``rotational parallax'')
of $730\pm168$ kpc, consistent with standard candle estimates.  Significant improvement
in the accuracy of this distance estimate can come from better \HI\ data and
a longer time baseline for the proper motion measurements.

Absolute proper motions (with respect to background quasars) have been
measured for two Andromeda satellites: M33 and IC~10.   Assuming these galaxies
are gravitationally bound to Andromeda, their 3-dimensional motions yield a
lower mass limit for Andromeda of $>7.5\times10^{11}$ \Msun\ (\citealt{Brunthaler:07}).
Of course, knowledge of the proper motion of Andromeda would refine this estimate
and is key to unlocking the history and fate of the Local Group.  Observations
with the upgraded bandwidth of the VLBA and the Green Bank and Effelsberg 100-m
telescopes are underway to measure the
proper motion of M~31* (the weak AGN at the center of Andromeda) and the
motions of a handful of \hho\ masers recently discovered in the galaxy 
(\citealt{Darling:11}).  Awaiting a direct measurement of the absolute proper motion 
of Andromeda, one can use the fact that M33 has not been tidally disrupted by a close
encounter with Andromeda in the past to place constraints on the motion of Andromeda.
In this way \citet{Loeb:05} showed that Andromeda likely has a proper motion of
$\sim100$ \kms.

\subsection{Tests of General Relativity}

The dominant uncertainty in modeling the Hulse-Taylor binary pulsar and 
measuring the effects of gravitational radiation on the binary orbit
comes from uncertainty in the accelerations of the Sun ($\To^2/\Ro$) 
and the pulsar ($\Theta(R)^2/R$) as they orbit the center of the Galaxy (\citealt{Damour:91}).  
Using recently improved values for the fundamental Galactic parameters reduces the 
uncertainty in the binary orbital decay expected from 
gravitational radiation by nearly a factor of four compared to using 
the IAU recommended values (\citealt{Reid:14}).  The dominant uncertainty
in the general relativistic test parameter is now dominated by the
uncertainty in the pulsar distance, and a VLBI parallax for the binary accurate 
to $\pm14$\% would bring the contribution from distance uncertainty down
to that of Galactic parameter uncertainty.

\citet{Deller09a} obtained an accurate parallax distance of $1.15^{+0.22}_{-0.16}$ kpc 
for the pulsar binary J0737$-$3039 A/B.  
Given that these pulsars are only a kpc distant (not $\sim10$ kpc as is the 
Hulse-Taylor pulsar), with this parallax accuracy uncertainties in
the correction for the effects of Galactic accelerations are 
an order of magnitude smaller than for the Hulse-Taylor system.  
Thus, with perhaps another decade of pulsar timing, one might achieve a
test of the effects of gravitational radiation predicted by general
relativity at 0.01\% level.  

Pulsars orbiting the super-massive black hole at the Galactic center,
Sgr~A*, may show general relativistic effects that can be
measured and tested based on high-accuracy astrometric observations.
Pulsars near the Galactic center are strongly
affected by interstellar scattering, which makes it
difficult to observe these pulsars at $\lax10$ GHz.
For instance, the recently-discovered transient magnetar PSR J1745$-$2900
(\citealt{Mori13}), located only 3'' from Sgr A*, 
is highly likely to be in the Galactic center region.
If another pulsar is found closer to Sgr~A*,
it could prove to be an excellent target for testing
general relativity with unprecedented accuracy.

Gravity Probe B (GP-B) was a satellite mission to test
general relativistic frame-dragging 
(Lense--Thirring effect) caused by the spin of the Earth.
The satellite was equipped with four ultra-stable gyroscopes, allowing
satellite altitude variation to be measured with unprecedented accuracy.
Based on the data from the four gyroscopes spanning $\sim$ 1 year, 
\citet{Everitt11} reported a geodetic drift at $-6601.8\pm 18.3$ \masy\ 
and a frame-dragging drift at $-37.2\pm7.2$ \masy.  These results are
fully consistent with the predictions of Einstein's theory of general
relativity ($-6606.1$ \masy\ and $-39.2$ \masy, respectively).
VLBI astrometry played a fundamental role in the
calibration of the GP-B data. IM Peg (HR~8703), an RS CVn type binary system, 
was used as the reference star.  Its parallax and proper motion were
accurately measured with VLBA observations spanning 5 years
(\citet{Shapiro12} and references therein).
Based on the astrometry of IM Peg relative to 3C454.3,
the absolute proper motion of IM Peg was determined with an
accuracy better than 0.1 \masy\ (\citealt{Ratner12}), which
provided the fundamental basis for measuring the geodetic and frame
dragging effect by GP-B.

The classical test of general relativity, measuring gravitational bending of
star light by the Sun (\citealt{Dyson:19}), has been repeated with radio 
interferometry by observing background radio sources and has provided one
of the most accurate tests of general relativity.
The most recent analysis of VLBI data reported that the
post-Newtonian parameter $\gamma$ is consistent with unity 
to an accuracy of one part in $10^{-4}$ (\citealt{Lambert09,Fomalont09,Lambert11}).
Additionally, \citet{Fomalont03} detected the deflection of radio waves
by Jupiter, including the retarded delay caused by Jupiter's motion, and 
claimed that this constrained the speed of gravitational waves, although 
this claim is controversial (see \citet{Asada02,Will03,Fomalont:09b}).

\subsection {Extrasolar Planets and Brown Dwarfs}

A radio astrometric search for extrasolar planets was conducted by 
\citet{Lestrade96} toward $\sigma^2$ CrB over a 7.5 year period.  
After removing the effects of parallax 
and proper motion, their post-fit residuals of $<0.3$ mas excluded a Jupiter-mass 
planet orbiting at a radius of 4 AU from the central star.  
\citet{Guirado97} observed the active K0 star AB Dor with 
the Australian LBA and, combined with HIPPARCOS data, inferred a companion 
with a mass of $\sim$0.1$M_\odot$.  
These studies demonstrate the potential of radio astrometry to detect
low-mass companions, including brown dwarfs and planets.

Low-mass stars ($<1$ \Msun) could have habitable planets which are difficult 
to detect with optical radial-velocity techniques, because the stars are 
faint and activity can distort their line profiles, reducing the accuracy of 
velocity measurements.  
Thus, radio astrometric planet searches are complementary to other approaches.  
The Radio Interferometric Planet (RIPL) Search (\citealt{Bower:09}) and
the Radio Interferometric Survey of Active Red Dwarfs (\citealt{Gawronski:13})
seek to detect the effects of planets around nearby low-mass stars,
specifically active M dwarfs.  For RIPL, the demonstrated astrometric
accuracy is 260~\uas, which for GJ~897A limits a planetary companion at 2~AU
radius to $<0.15~{\rm M}_J$ (\citealt{Bower:11}).

\subsection {AGN Cores}

Active Galactic Nuclei (AGNs) often show very strong and compact radio
emission, with brightness temperatures up to and exceeding 10$^{12}$ K.
\citet{Bartel86} measured the relative positions of a radio bright pair of QSOs,
NRAO 512 and 3C 345, with a global VLBI network, revealing that the core positions 
of the sources were stable to within $\pm20$ \uasy.
This stability corresponds to an upper limit of $\sim1 c$ for the core-motions,
which is significantly lower than observed in the jets of super-luminal sources.
\citet{Marcaide94}, \citet{Rioja97}, \citet{Guirado95} and \citet{Marti-Vidal08}
reported little if any relative proper motion between other QSO pairs (at levels
of $\sim10$ \uasy), consistent with contamination of stationary cores with
some expanding plasma.  Such studies point out a source of limiting ``noise'' 
when using radio loud QSOs for establishing fundamental reference frames.

\begin{figure}[h]
  \center{\includegraphics[width=0.8\textwidth]
  {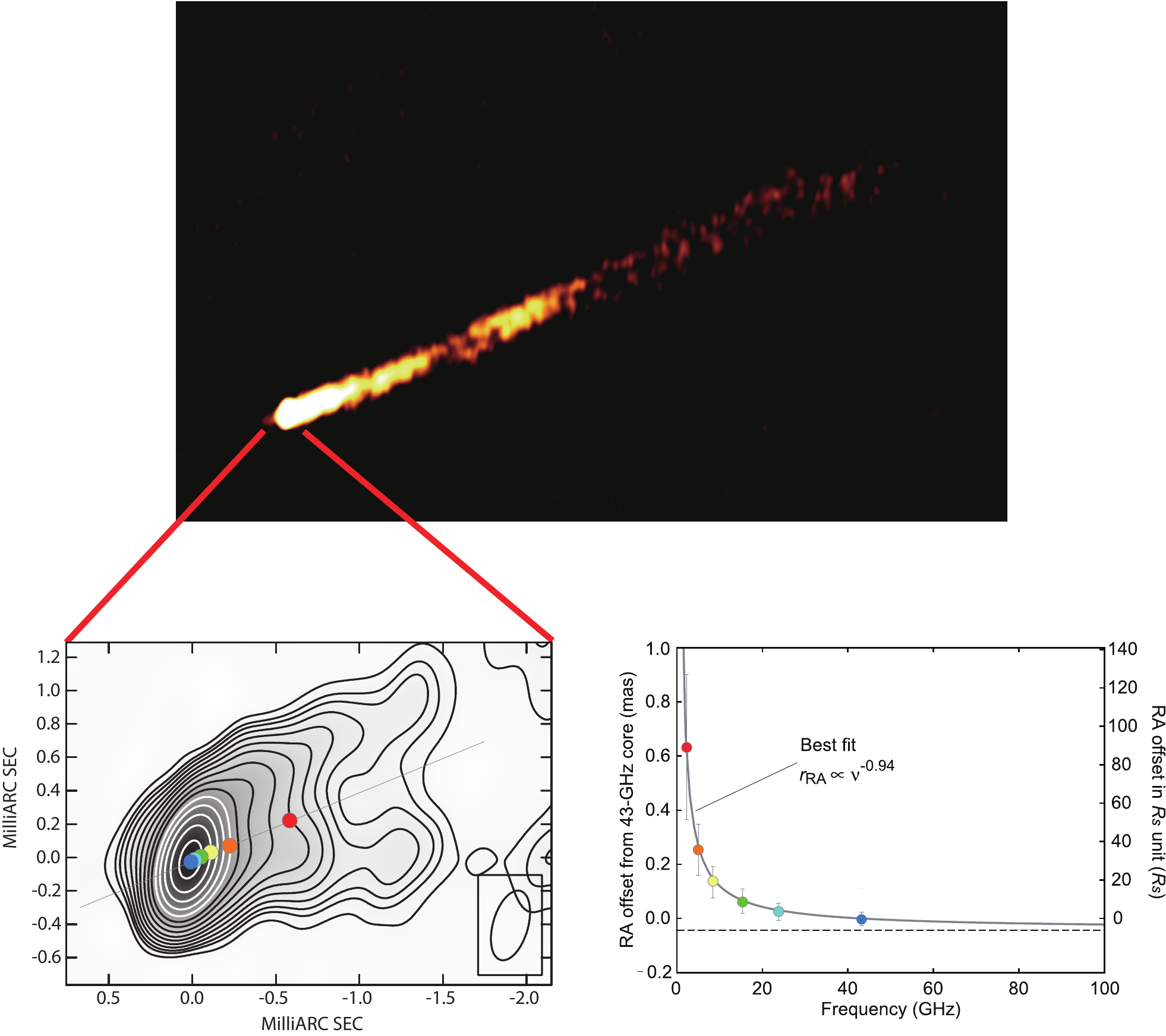}}
  \caption{The apparent shift in the core of M 87 as a function of
           observing frequency after \citet{Hada12}.  This effect
           can be explained by frequency dependent optical depths
           and allows the position of the supermassive black hole to
           be accurately located on the images.}
 \label{fig:M87}
\end{figure}

In addition to measuring relative positions between two QSOs, one can
precisely measure positions of jet components within a source at different 
frequencies.  In this manner, \citet{Hada11} measured the 
theoretically predicted core-shift as a function of observing frequency
(\citealt{Blandford:79}) for the super-massive black hole of M~87
(see {\bf Figure~\ref{fig:M87}}) and located the black hole relative 
to the jet emission with an accuracy of $\sim10 R_{\rm sch}$.
\citet{Kovalev08} and \citet{Sokolovsky11} have confirmed that
frequency-dependent core-shifts, such as observed in M~87, are
common in large samples of AGNs with prominent jets.  
\citet{Bietenholz04} used SN 1993J in M~81 as a position reference for 
multifrequency imaging of M~81* (the AGN in that galaxy) to accurately 
locate the super-massive black hole.  Later \citet{Marti-Vidal11} 
found that the M~81* core position appears variable at lower frequencies, and they
suggest this is a result of jet precession.

\citet{Porcas09} considered the effects of an AGN core-shifts on
group-delay astrometry, used for measuring antenna locations for
{\it geodetic} VLBI and for establishing reference frames (\eg\ the ICRF). 
Usually group-delay measurements are conducted at 8.4 GHz and, to remove 
propagation delays in the ionosphere, simultaneous observations are also  
made at 2.3 GHz.  When removing the dispersive component of delay to
correct for ionospheric propagation delay, generally one assumes that AGN core 
positions are the same at the two observing frequencies.
However, the core-shift effect on dual-frequency geodetic
VLBI observations can be comparable to a position error of $\sim170$ \uas\
and should be measured and compensated for in the future, for example, 
when tying the radio reference frame with a new optical frame constructed 
by Gaia (\citealt{Bourda:11}).

\subsection {Satellite Tracking}

The technique of high-accuracy VLBI astrometry can be applied to locating
spacecraft, using telemetry signals as radio beacons.
In fact, astrometric observations of radio beacons placed on the moon by 
Apollo missions were among the earliest applications of phase-referencing VLBI.
The Apollo Lunar Surface Experiments Package and Lunar Roving Vehicle
radio transmitters from various Apollo missions spanned hundreds of km 
across the Lunar surface.  These transmitters were observed by
\citet{Counselman73} and \citet{Salzberg73}, who measured their separations 
with an accuracy of $\sim1$ m ($\sim0.5$ mas).
Later \citet{Slade77} was able to tie these transmitters to the celestial
reference frame with a positional accuracy of $\sim$1 mas.   \citet{King76}
combined VLBI observations of Lunar transmitters
with ranging data to better estimate the mass of the Earth-Moon system 
and the moment-of-inertia ratios of the moon.  (See \citet{Lanyi07} for a
recent review.)

Recently, high-accuracy radio astrometry played a part in
generating a precise map of the Lunar gravity field in conjunction
with the SELENE mission (\citealt{Goossens11}).  This was made possible by the 
high relative-position accuracy possible with in-beam astrometry (\citealt{Kikuchi09}).
This method was originally developed in 1990's for observing
artificial radio sources near Venus (\citealt{Border92,Folkner93}) and
made possible measurements of wind speed as a function of altitude, by 
monitoring the trajectories of probes released into the Venusian atmosphere 
(\citealt{Counselman79,Sagdeyev92}). 

In the outer solar system, VLBI astrometry played a role in the
PHOBOS-2 (\citealt{Hildebrand94}) and Mars Odyssey missions (\citealt{Antreasian05}).
The motion of the Huygens probe as it descended in Titan's atmosphere 
revealed the vertical profile of its wind speed (\citealt{Witasse06}).   
The Cassini satellite itself was used to trace 
the gravity field of Saturn, and \citet{Jacobson06} combined VLBI data with optical
observations and Doppler ranging to better constrain the mass and 
potential of the Saturnian system.  Also, VLBA observations of Cassini
have measured the center-of-mass of the Saturnian system with an accuracy 
of 2 km (0.3 mas) with respect to the ICRF (\citealt{Jones11}).  
The potential of radio astrometry for improving future space missions has been 
reviewed by \citet{Duev12}.

\section{Arrays for Radio Astrometry} \label{sect:arrays}

There are four major VLBI arrays that are regularly doing radio
astrometry: the VLBA (Very Long Baseline Array) in US, the EVN (European
VLBI Network), the VERA (VLBI Exploration of Radio Astrometry) array in Japan, 
and the LBA (Long Baseline Array) in Australia.  Basic parameters of these 
arrays are summarized in {\bf Table \ref{table:VLBI-array}}.

\begin{table}
\footnotesize
\def~{\hphantom{0}}
\caption{Major Radio Astrometric Interferometers}
\label{table:VLBI-array}
\hspace{-0cm}\begin{tabular}{@{}llllll@{}}
\toprule
Array & Country/ & Antenna diameters \&  & Maximum & Operating &
	      beam size (mas) \\
      & region   & number in array & baseline (km) & frequencies &   \\
      &  & & & & comments \\
\colrule
VLBA & USA & 25m$\times$10 & 8600 km & 0.3--86 GHz & 0.17 at 43
 GHz \\
 & \multicolumn{5}{r}{homogeneous and best imaging capability}\\
\hline 
VERA & Japan & 20m$\times$4 & 2300 km & 6.7--43 GHz & 0.63 at 43 GHz \\
 & \multicolumn{5}{r}{dedicated to astrometry with dual-beam system}\\
\hline 
EVN  & Europe & 14m--100m, $\times \sim10$ & 3000-10000 km &
 1.6--22 GHz & 0.30 at 22 GHz \\
 & \multicolumn{5}{r}{high sensitivity with large dishes}\\
\hline 
LBA  & Australia & 22m--70m, $\times \sim 10$ & 1700 km & 1.4--22 GHz & 1.7
 at 22 GHz \\ 
 & \multicolumn{5}{r}{only VLBI array in the southern hemisphere}\\
\hline\hline
JVLA & USA & 25m$\times$27 & 36 km & 0.07--50 GHz & 40 at 43 GHz \\ 
 & \multicolumn{5}{r}{ connected array with high-sensitivity and excellent imaging}\\
\hline 
ALMA & Chile & 12m$\times$54 + 7m$\times$12 & 16 km & 43--900 GHz & 4.5
at 850 GHz \\ 
 & \multicolumn{5}{r}{large connected array at mm and sub-mm wavelengths}\\
\hline 
SKA  & Au/SA & $\sim$1 km$^2$ aperture & $\sim 3000$ km &
 0.1--22 GHz & 0.7 at 22 GHz\\ 
 & \multicolumn{5}{r}{future large cm-wavelength array}\\
\botrule
\end{tabular}
\end{table}

The VLBA consists of ten 25-m diameter 
telescopes spread over US territory from Hawaii in the west to 
New Hampshire and Saint Croix in the east.
With a maximum baseline length of $\approx8000$ km and frequency coverage
from 300 MHz to 86 GHz, the VLBA is a flexible and sensitive array.
The homogeneity of the array, with all the antennas and instruments
identical, is advantageous as it makes calibration straightforward.
While the VLBA is a general-purpose imaging array, recently
phase-referencing astrometry has occupied a major portion of its observing time.
Currently there are several large astrometric projects on the VLBA.
The Bar and Spiral Structure Legacy (BeSSeL) Survey is measuring
parallaxes and proper motions of hundreds of Galactic masers associated
with high-mass star forming regions.
The Gould's Belt Distances Survey aims to measure distances to and provide
a detailed view of star-formation in the Solar neighborhood.
The Radio Interferometric Planet Search (RIPL) is an astrometric search 
for planets around nearby low-mass stars.
The Pleiades Distance Project seeks absolute parallaxes for up to 10
cluster stars in order to resolve the current cluster distance controversy
and provide a solid foundation for many aspects of stellar astrophysics. 
Finally, the Megamaser Cosmology Project seeks to measure \Ho\ directly 
for 10 \hho\ megamaser galaxies well into the Hubble flow in order to
independently constrain \Ho\ with $\pm3$\% accuracy. 

\begin{figure}[h]
  \center{\includegraphics[width=0.7\textwidth]
          {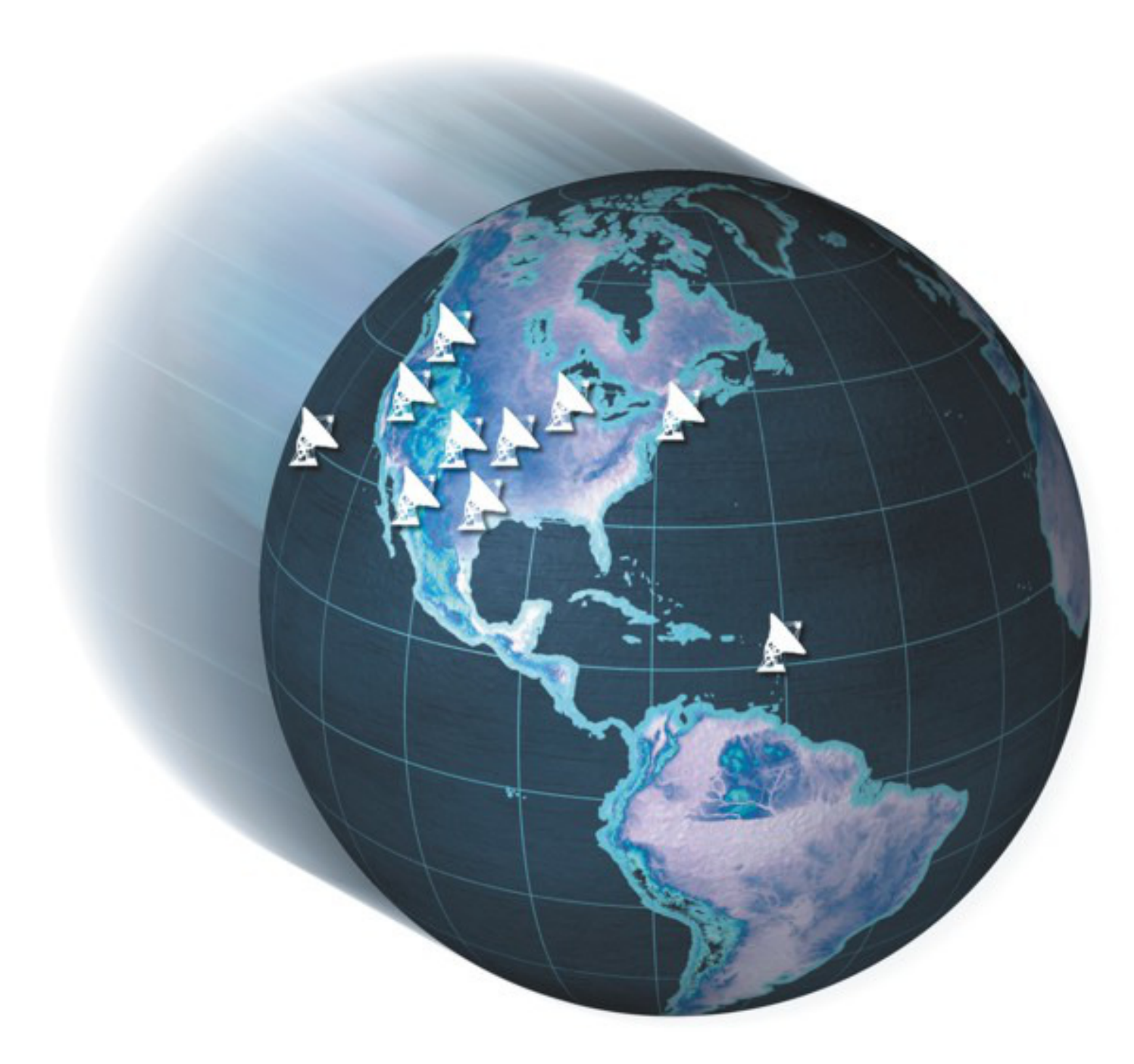}
         }
  \caption{Schematic portrayal of the locations of the Very Long Baseline Array 
          (VLBA) antennas.  The VLBA has ten 25-m diameter antennas spanning
          the globe from Hawaii to New Hampshire and St. Croix.  
          }
 \label{fig:VLBA}
\end{figure}

\begin{figure}[h]
  \center{\includegraphics[width=0.7\textwidth]
          {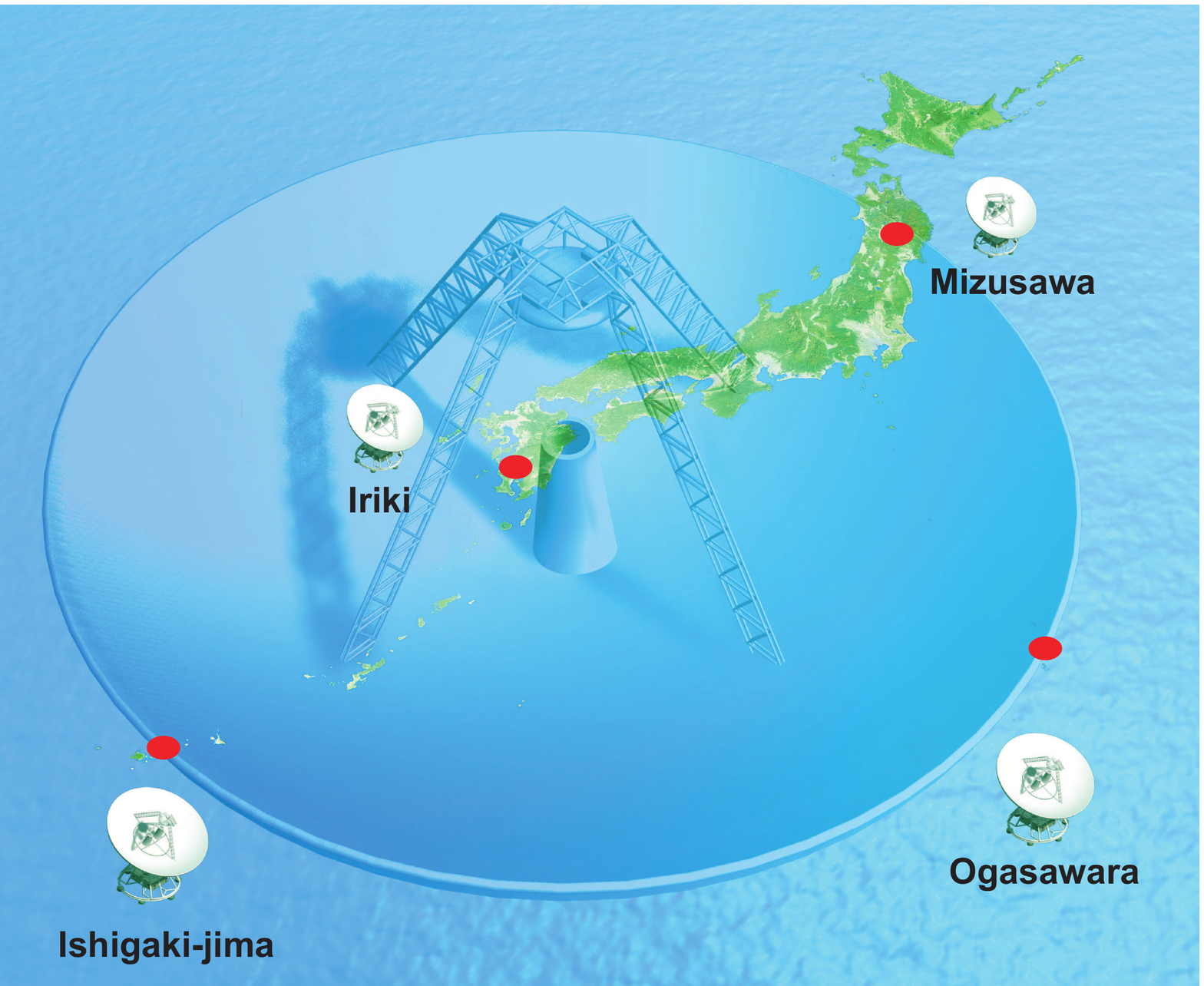}
         }
  \caption{Schematic portrayal of the locations of the VERA antennas.  
           The VERA array has four 20-m diameter antennas spanning
           the Japan and is dedicated to astrometric observations.
          }
 \label{fig:VERA}
\end{figure}

The VERA array consists of four 20-m diameter telescopes located across Japan 
with a maximum baseline length of 2300 km (see {\bf Figure~\ref{fig:VERA}}).
Each antenna is equipped with two receiver systems that are independently steerable 
in the focal plane.  As such VERA can observe target and reference sources 
simultaneously to effectively cancel tropospheric fluctuations.
VERA is also the only array dedicated full-time to phase-referencing astrometry.
Most of the observing time is spent on parallax measurements of maser sources 
tracing spiral structure in the Milky Way and of red giant stars.

The EVN array has antennas distributed across Europe as well
as in other countries, including China, South Africa and USA.
The EVN is most sensitive at frequencies $<10$ GHz, owing to some 
large antennas: the Effelsberg 100-m, the Jodrell Bank 76-m and soon 
the Sardinia 64-m telescope.
The array has been used for astrometric measurements of OH masers at 
1.6 GHz, methanol masers at 6.7 GHz, and active stars.

The LBA in Australia is the only VLBI array regularly operating in the
southern hemisphere.  It has high sensitivity when the Parkes 64-m and
Tidbinbilla 70-m telescopes are included.
The LBA has provided astrometric measurements for southern pulsars,
and hopefully it will soon be used for maser parallaxes.
This is necessary in order to trace the 3-dimensional structure of the 
roughly one-third of the Milky Way that cannot be observed from the north.

\section{Fundamentals of Radio Astrometry}

\begin{figure}[h]
  \center{\includegraphics[width=0.8\textwidth]
  {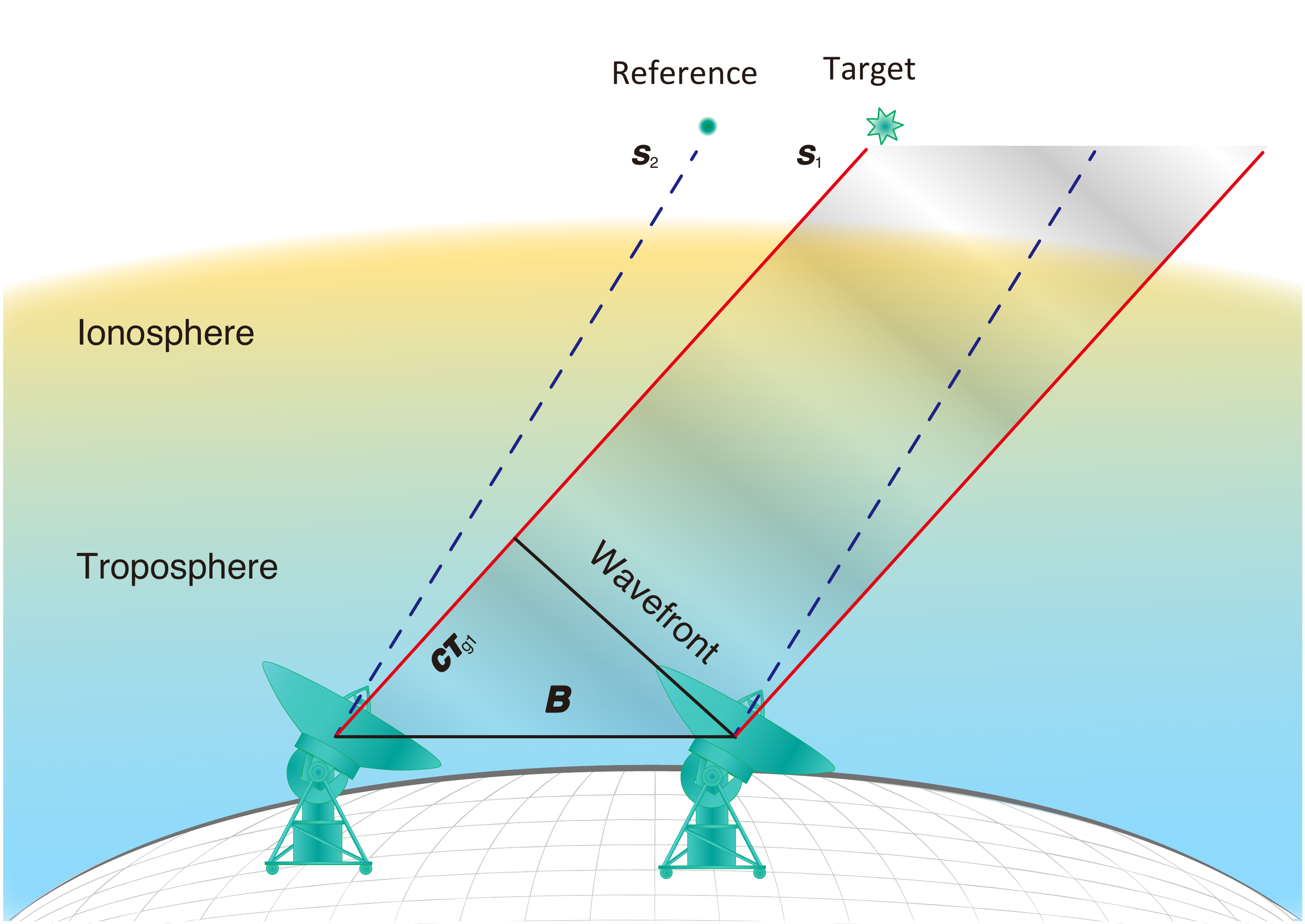}}
  \caption{Schematic view of a delay measurement in phase-referencing
 astrometry. Here, for simplicity, an array with two stations is shown.
 The baseline vector and source directions are indicated by lines. 
 In relative astrometry, the target and adjacent
 reference sources are observed (nearly) at the same time so 
 delay errors can be effectively canceled in the relative measurement.
 For relative radio astrometry, the dominant error sources are generally 
 uncompensated propagation delays in troposphere and/or ionosphere.}
 \label{fig:geometry}
\end{figure}

The fundamental observable for a radio interferometer is the arrival time
difference of wavefronts between antennas, owing to the finite 
propagation speed of electro-magnetic waves (the speed of light $c$).
For an ideal interferometer, the arrival time difference is determined
by the locations of the antennas with respect to the line-of-sight to the source and is
referred to as the ``geometric delay.''  The basic equation of a radio interferometer, 
which relates the geometric delay, $\tau_g$, to the source unit vector, $\vec{s}$, 
and the baseline vector,  $\vec{B}$, is given by
\begin{equation}
\label{eq:geometric-delay}
 \tau_g = \frac{\vec{s}\cdot\vec{B}}{c}~~.
\end{equation}
In {\bf Figure~\ref{fig:geometry}}, we sketch the geometry of these vectors.
Note that while the source vector $\vec{s}$ has two parameters (\eg\ right ascension
and declination), the geometric delay is a scalar and so multiple measurements of 
geometric delays are required to solve for a source position.
Such measurements can be made, even with a single baseline, by utilizing the Earth's 
rotation, which changes the orientation of the baseline vector with respect to the 
celestial frame.

We now consider the effects of observational errors on astrometric accuracy.
Given the uncertainty in a delay measurement, which we denote 
$\Delta \tau$, from Equation \ref{eq:geometric-delay} one can roughly
estimate the astrometric error, $\Delta s$: 
\begin{equation}
\label{eq:delay-error}
 \Delta s \approx \frac{c \Delta \tau}{|B|}~~.
\end{equation}
As seen in Equation \ref{eq:delay-error}, for a given $\Delta \tau$,
astrometric accuracy improves with longer baselines.   
This is the fundamental reason why VLBI,
which utilizes continental and/or inter-continental baselines, 
can achieve the highest accuracy astrometry.
For example, for an 8,000 km baseline and a typical delay error of 
$\sim$2 cm (converted to path length by $c\Delta\tau$),
one expects an astrometric error of $\sim0.5$ mas.
This is a typical error for {\it absolute} astrometry, such as
measuring QSO positions in the ICRF using broad-band delay measurements.

For Galaxy-scale parallaxes, however, this accuracy is insufficient as
parallaxes can be $\sim0.1$ mas, corresponding to distances of $\sim10$ kpc.
At such distances, one needs astrometric accuracy of $\pm10$ \uas\ 
to achieve $\pm10$\% uncertainty.  In order to obtain \uas-level accuracy,  
one must substantially cancel systematic delay errors through
{\it relative} astrometry, \ie\ a position measurement with respect to a nearby
reference source.  Historically, the concept of relative astrometry has
been attributed to Galileo, who proposed measuring the parallax of a nearby
star with respect to a more distant star located close by on the sky.

\subsection{Relative Astrometry} \label{sect:relative}

The delay measured with an interferometer can be considered as the sum of the 
geometric delay (which we would like to know) and additional terms (which need 
to removed from the data):
\begin{equation}
\label{eq:tau-obs}
 \tau_{\rm obs} = \tau_{\rm g} + \tau_{\rm tropo} + \tau_{\rm iono} +
  \tau_{\rm ant} + \tau_{\rm inst} + \tau_{\rm struc} +\tau_{\rm therm}~~.
\end{equation}
Here $\tau_{\rm tropo}$ and  $\tau_{\rm iono}$ are the delays in the propagation 
of the signal through the troposphere and ionosphere, respectively,
$\tau_{\rm ant}$ is caused by an error in the location of 
an antenna, $\tau_{\rm inst}$ is an instrumental delay in the telescope or
electronics, $\tau_{\rm struc}$ is the delay due to unmodeled source structure, 
and $\tau_{\rm therm}$ is the uncertainty in measuring delay caused by thermal noise.
Note that generally throughout this review the terms {``}delay{''} and
{``}phase{''} (or phase-delay) can be used interchangeably:
for instance, the interferometric phase caused by the geometric delay in
Equation \ref{eq:geometric-delay} can be written as,
\begin{equation}
\label{eq:phase-delay}
 \phi_{\rm g} = 2\pi \nu \tau_{\rm g}~~,
\end{equation}
where $\nu$ is the observing frequency.

For relative astrometry, one observes two sources, the target and
reference, at nearly the same time and nearly the same position on the sky, 
and differences the observed delays (phases) between the pair of sources.
This type of observation is often referred to as ``phase-referencing,'' 
as the observed phase of the reference source is used to correct the phase of 
the target.  Applying Equation~\ref{eq:tau-obs} for the difference between
two sources, we obtain the delay difference observable:
\begin{eqnarray}
 \label{eq:delta-tau-obs}
 \Delta\tau_{\rm obs} &=& (\tau_{\rm geo,1} - \tau_{\rm geo,2}) \nonumber \\
                   &+& (\tau_{\rm tropo,1} - \tau_{\rm tropo,2})
                    +  (\tau_{\rm iono,1} - \tau_{\rm iono,2}) \nonumber \\
                   &+& (\tau_{\rm ant,1} - \tau_{\rm ant,2})
                    + (\tau_{\rm inst,1} - \tau_{\rm inst,2}) \nonumber \\
                   &+& (\tau_{\rm struc,1}- \tau_{\rm struc,2})
                    + (\tau_{\rm therm,1} - \tau_{\rm therm,2})~~.
\end{eqnarray}
Here subscripts 1 and 2 denote the quantities for the target and
reference source, respectively.
The first term is the difference in geometric delay between
the source and reference, which corresponds to the relative position of the
target source with respect to the reference source.
The additional terms in the Equation \ref{eq:delta-tau-obs} may be
reduced if they are similar for the two lines of sight. 
As will be discussed in detail later, four terms ($\tau_{\rm trop}$,
$\tau_{\rm iono}$, $\tau_{\rm star}$, and $\tau_{\rm inst}$) are
antenna-based quantities (delays originated at each antenna) and generally
are similar for the target and reference lines of sight, for small source
separations.

Antenna-based terms can be effectively reduced by phase-referencing.  
For example, the delay error generated by an antenna 
position offset, $\Delta \vec{B}$, is given by
\begin{equation}
\label{eq:delay-ant}
\tau_{\rm ant} = \frac{\vec{s}\cdot \Delta \vec{B}}{c} \sim \frac{|\Delta B|}{c}~~. 
\end{equation}
Note the approximation used to obtain the right hand side of Equation~\ref{eq:delay-ant}
is that the trigonometric terms in the vector dot product are generally $\lax1$.
Differencing the antenna delays for two adjacent sources, yields
\begin{equation}
\label{eq:delta-delay-ant}
\Delta \tau_{\rm ant} = (\tau_{\rm ant,1} - \tau_{\rm ant,2}) 
  = \frac{(\vec{s_1}-\vec{s_2})\cdot \Delta \vec{B}}{c} \sim \theta_{\rm sep}
\frac{|\Delta B|}{c}~~,
\end{equation}
where $\theta_{\rm sep}$ is the separation angle between the sources. 
Comparison of Equations \ref{eq:delay-ant} and \ref{eq:delta-delay-ant} shows
that the delay error caused by an antenna position error is reduced by a
factor of $\theta_{\rm sep}$ in the differenced delay.
This reduction can significantly improve {\it relative} over {\it absolute} 
astrometric accuracy.  For a separation of $\theta_{\rm sep}=1^\circ$, 
the error reduction factor ($\theta_{\rm sep}$) is about 0.02 (radians).

As we will discuss later in more detail, the dominant error source in
radio astrometry is uncompensated propagation delays, generally tropospheric
for observing frequencies $\gax10$ GHz and ionosphere for lower frequencies.
These terms are ``antenna-based,'' just as the antenna location errors described above.
A rough estimate of astrometric accuracy achievable in the presence of
propagation delay errors is given by
\begin{equation}
\label{eq:delay-error-difference}
  \Delta s_{\rm rel} \approx \theta_{\rm sep} \frac{c \Delta \tau}{|B|}~~.
\end{equation}
The reduction in position uncertainty in going from Equation \ref{eq:delay-error}
to Equation \ref{eq:delay-error-difference} is, of course,
the addition of the canceling term, $\theta_{\rm sep}$.
Using the same example values previously used for absolute astrometry,
($|B|$=8000 km and a delay error of $c\Delta t \sim 2$ cm), relative
astrometric accuracy for a $1^\circ$ source separation is $\sim10$ \uas.

\subsection{Sources of Delay Error}

Uncompensated delay differences between antennas in an interferometric array 
usually limit radio astrometric accuracy.  In this subsection, we discuss
sources of delay errors individually.  

\subsubsection{Troposphere} \label{sect:troposphere}

The propagation speed of electro-magnetic waves in the troposphere is slower than 
in vacuum, causing an extra delay ($\tau_{tropo}$) to be added to the geometric delay.
The tropospheric delay is almost entirely non-dispersive (frequency independent)
at radio frequencies.
It is convenient to separate the tropospheric delay into two components based
on the timescales of fluctuation: a slowly ($\sim$ hours) and rapidly ($\sim$ minutes)
varying term.

The rapidly varying term is associated with the passage of small ``clouds'' 
of water vapor flowing at wind speeds of $\sim 10$ m s$^{-1}$ 
at a characteristic height of $\sim1$ km over each antenna. 
This can be the main cause of coherence loss in interferometric observations.
The characteristic time scales for interferometer phase fluctuations induced by 
tropospheric water vapor can be described by the Allan standard 
deviation ($\sigma_{\rm A}$) and is typically $\sim0.5~{\rm to}~1\times10^{-13}$ over 
timescales of 1 to 100 sec (\citealt{TMS01,Honma:03}).
An interferometer coherence time, $\tau_{\rm coh}$, can be defined as the time interval 
over which phase fluctuation differences between two antennas
accumulate one radian.
At an observing frequency $\nu$, this implies
\begin{equation}
 2\pi \sigma_{\rm A} \nu \tau_{\rm coh} \sim 1~~.
\end{equation}
For $\sigma_{\rm A}=0.7\times10^{-13}$ and $\nu=22$ GHz, this implies a coherence time of 
$\sim100$ sec.  
One must measure and remove the phase variations on a time-scale shorter than the 
coherence time.  Rapid switching between the target and calibrator
sources can usually accomplish this, provided the antennas can slew and
settle on sources rapidly (see \S\ref{sect:methods}).

The rapid variations from tropospheric delays can be completely removed if one 
can simultaneously observe the target and calibrate sources .  
This is done with the VERA antennas, which have dual-beam observing systems.
Alternatively, this can also be accomplished if one is fortunate in finding a 
source pair with a very small angular separation, so that both fit within the primary 
beam of the antennas (``in-beam'' calibration).
Note that with either method, this still leaves a spatial variation component, 
but that term is generally smaller than the temporal term for source separations 
of a few degrees.

The slowly varying delay term is generally associated with the ``dry'' part of the 
troposphere, although the slow term can also contain a small, relatively stable 
contribution from water vapor.  At sea-level, a typical path delay for the 
dry component is 230 cm at the zenith and water vapor can contribute 
up to a several tens of cm for very humid locations.
The total dry delay can be estimated from antenna latitude and elevation,
and if necessary using local temperature and pressure values.
Since the total dry delay is considerable, it is important to remove 
its effects during correlation, or early in post-correlation processing,
by carefully accounting for the exact slant path through the atmosphere
(air mass), which will be different for the target and calibrator.
The slowly varying wet component cannot reliably be estimated from surface
weather parameters and needs to be directly measured and removed to achieve
\uas\ astrometry (see \S\ref{sect:geoblocks} and \S\ref{sect:GPS} for details).

\subsubsection{Ionospheric delay}

The ionosphere is a partially ionized layer located between 
$\sim50$ and $\sim500$ km altitude.  
An electromagnetic wave propagating through this plasma experiences
phase and group delays that are frequency dependent (\ie\
dispersive delays).  The phase and group delay are given by
\begin{equation}
\tau_{\rm iono} \equiv {1\over{2\pi}} {\phi\over\nu} = - \frac{cr_0}{2\pi \nu^2} I_e~~,
\end{equation}
and
\begin{equation}
\tau_{\rm grp, iono} \equiv {1\over{2\pi}} {\partial\phi\over\partial\nu} = \frac{cr_0}{2\pi \nu^2} I_e = - \tau_{\rm iono}~~,
\end{equation}
where $r_0$ is the classical electron radius and $I_e=\int n_e dl$ is
the electron column density along the line of sight or total electron content (TEC).
Due to the dispersive nature of plasma, delays 
are larger at lower frequency because of the $\nu^{-2}$ dependence.
(Note that the ionospheric phase delay has the opposite sign of the group delay;
therefore correction of interferometer phase for ionospheric delays is
different than for tropospheric delays.)

The ionization of the ionosphere is predominantly from solar
radiation, and thus there is a strong diurnal variation, as well as from
the 11 year solar cycle.
Due to the diurnal variation, the ionospheric delay needs to be
modeled (or measured) and calibrated on hour scales.
Modeling of the ionospheric delays is done by a combination
of a vertical TEC (VTEC) and a mapping (air mass like) function.
Typical values for VTEC range from a few to 100 TECU, where 1 TECU 
corresponds to an electron column density of $10^{16}$ m$^{-2}$.
Once the TEC toward a source is known, the ionospheric delay (in path
length units) can be calculated by the following relation,
\begin{equation}
 |c\tau_{\rm iono}| = 40.3 \left(\frac{I_e}{{\rm TECU}}\right)
  \left(\frac{\nu}{{\rm GHz}}  \right)^{-2} \;\;\; {\rm (cm)}~~.
\end{equation}
At a frequency of 22 GHz, a 50 TECU column density causes a delay
equivalent to an extra path length of $\sim4$ cm, but it reaches $\sim400$ cm 
at a frequency of 2 GHz.

Because of the dispersive nature of the ionospheric delay,
dual-frequency observations are effective for calibration.
Most geodetic measurements involve simultaneous observations at
2.3 and 8.4 GHz in order to calculate and then remove ionospheric delays.
Another way to measure ionospheric TEC values is to use GPS data, which 
provide dual-frequency signals at 1.23 and 1.58 GHz.
Several services accumulate GPS data from stations around the Earth
and provide global TEC models approximately every two hours.  Generally,
radio interferometric data are calibrated for ionospheric delays
using these models.

\subsubsection{Instrumental Delay}

Radio propagation in antenna structures, feeds, and electronics causes additional
delays, which we lump together as ``instrumental delays'' ($\tau_{inst}$).
In addition, there are electronic phase offsets associated with the
generation of the local oscillators used to mix the observed frequencies
to ``intermediate'' frequencies ($\sim1$ GHz) prior to correlation.
The data recorded at each antenna are time-tagged using a clock tied to
the fundamental frequency standard at each station 
(usually a hydrogen maser atomic oscillator),
and owing to slight frequency offsets (that do not affect local oscillator 
stability) these clocks generally gain or lose time at a rate of one part in 
$\sim10^{-14}$ (only $\sim1$ nsec per day!).  
However, a delay error of 1 nsec causes a phase
slope of $\pi$ radians across a 500 MHz observational bandwidth, and it is
critical to correct for this.  Calibration observations using ``geodetic blocks''
(see \S\ref{sect:geoblocks}) can remove clock errors to an accuracy of $\lax0.1$ nsec.
Other instrumental delay and phase offsets are generally relatively slowly varying,
for well designed systems and temperature controlled electronics, and can be
calibrated and removed by observations of strong sources a few times a day.

Any residual instrumental delay and phase offsets will be mostly canceled
when switching between two sources, provided the electronics are not changed. 
Note that if one of the sources is a spectral line and the other a continuum
emitter, one should try to place the line near the center of the continuum band
to ensure better calibration.  On the other hand, if the two sources are observed 
with different receivers, which is the case for dual-beam receiving at VERA,
there is a need for additional calibration of the instrumental delay and phase, 
such as by using an artificial noise source (see \S \ref{sect:methods}).

\subsubsection {Antenna Position}

The antenna position is usually defined as the intersection 
of azimuth and elevation axes and can be measured to an accuracy 
of $\lax3$ mm with regular geodetic observations.  
The Earth's rotation rate varies (measured as a time correction, UT1$-$UTC) 
as does the location of an antenna on the Earth's crust with respect to 
its spin axis (polar motion).  Together these are called Earth 
Orientation Parameters (EOP).
EOP values are determined on a daily basis by the International Earth Rotation
and Reference Systems Service (IERS) by utilizing global
GPS data and geodetic VLBI observations.
A typical error in EOP values is $\approx0.1$ mas, provided
the final (not preliminary or extrapolated) values are used.
With high accuracy antenna locations and EOP corrections, the total 
uncertainty in the position of an antenna contributes less to the
astrometric error budget than uncertainty in atmospheric propagation delays.

\subsubsection{Source Structure}

If the observed sources are not point like, interferometric delay/phase can be 
affected by source structure.  Since structure phase shifts are independent 
for the target and reference sources, their effects do not cancel when
differencing the phases in relative astrometry.
Source structure phases are baseline-based quantities.  Because there are $\sim N^2$ 
baselines for $N$ antennas in an array, structure phase can be separated from 
station-based quantities and estimated using self-calibration (closure techniques 
and hybrid-mapping).
Once a source image is obtained, phase shifts caused by source structure can be 
calculated and subtracted from observed interferometric data.
Note that since self-calibration loses absolute position information, if one 
self-calibrates the reference source data, one must apply the {\it same} calibrations
to both the reference and target source to preserve relative astrometric accuracy.   
Generally it is not a good idea to self-calibrate the target source data, 
but instead measure the structure in the phase-referenced images.

\subsubsection{Thermal noise}

Thermal (usually dominated by receiver) noise is random and cannot be
calibrated and removed.
However, a random process does not cause systematic offsets, and by 
integrating over many samples its effects can be reduced.
Thermal noise in an interferometric image leads to position measurement 
uncertainty given by Equation~\ref{eq:geometric-delay} of \citet{Reid:88}:
\begin{equation}
\Delta\theta_{\rm therm} \approx 0.5 \frac{\theta_{\rm beam}}{SNR} 
  \approx 0.5 \frac{\lambda}{B}\frac{1}{SNR}~~.
\end{equation}
For a baseline length $B=8000$ km and observing wavelength $\lambda=1.3$ cm, 
the synthesized beam (FWHM) size is $\theta_{\rm beam}\approx0.3$ mas. 
Thus, for a source with an image signal-to-noise ratio $SNR\sim30$, the expected
position error due to thermal noise is only $\approx5$ \uas.
Therefore, thermal noise often is not a major source of radio astrometric
error.

\subsection{Observing methods} \label{sect:methods}

Several methods of phase-referencing are commonly used:
source switching, dual-beam, and in-beam observations.
Source switching or ``nodding'' is the most-commonly used observing method, 
because it does not require a special radio telescope and can be used for all 
source pairs. The only requirement is that slewing/settling times are short enough
to track tropospheric phase fluctuations.
As discussed in \S \ref{sect:troposphere}, coherence times at an observing
frequency of 22 GHz are $\sim100$ sec.  Therefore switching cycles should be 
shorter than this time, \eg\ a cycle of 60 sec obtained by integrating for
20 sec on the reference, 10 sec for slewing to the target, integrating for 20 sec 
on the target, then 10 sec for slewing back to the reference.
Because of the non-dispersive nature of tropospheric delays, the
interferometer coherence time scales linearly with wavelength,
i.e., $\tau_{\rm coh}\propto \lambda$.
Hence at a short wavelengths, the coherence time can be problematic. 
For an observing wavelength of $\lambda=1.3$ mm ($\nu=230$ GHz), 
the coherence time is likely to be $\sim10$ sec.  This makes it practically 
impossible to conduct switched VLBI observations at $\lax1$ mm wavelength.

In dual-beam observations, two sources are observed simultaneously using
independent feeds and receivers.  
This can be done using multi-feed systems on a single antenna, as with VERA,
or with multiple antennas at each site (\citealt{Rioja:09,Broderick:11}).
There is no gap between the observations of the target and reference sources, 
and hence there is no coherence loss owing to temporal phase fluctuations.  
The antennas of the VERA array achieve this with two
receivers installed at the Cassegrain focus.  These receivers are independently
steerable with a Steward-mount platform, and one can observe a source pair 
with separation angles between $0.3^\circ$ to $2.2^\circ$.
In dual-beam radio astrometry, special care needs to be taken to calibrate the 
instrumental delay, because it is not common to the target and reference source 
and thus will not cancel when phase-referencing.

In-beam observations can be regarded as a special case
of phase-referencing, where the target and reference sources are so closely
located that the two sources can be observed simultaneously with a single
feed (within the primary beam of each antenna).
In such a case, calibration error can be very effectively reduced, 
because the observations are done at the same time for the target and
reference, and, because the separation angle is small,
most systematic errors are largely canceled.   However, finding
a sufficiently strong reference source may be difficult; as such
in-beam observations are more common at lower observing frequencies, as the
primary beam becomes larger and the reference sources stronger.

\section{Advanced Techniques}

For parameters typical of cm-wavelength VLBI observations
seeking to measure relative positions between two sources separated by $\approx1^\circ$
with $\approx10$ \uas\ accuracy, phase reference sources should have coordinates 
known to $\lax5$ mas, antenna locations known to $\lax1$ cm, and electronic, 
clock and propagation delays known to $\lax0.05$ nsec.  

\subsection {Tropospheric Delay Calibration}

Tropospheric (non-dispersive) delays are usually calibrated by one of
two methods: 1) using geodetic-like observing blocks or 2) using GPS data.

\subsubsection {Geodetic Block Calibration:} \label{sect:geoblocks}

One can use observations of radio sources spread over the sky to measure 
broad-band (group) delays.  For sources with positions known to better than 
$\approx1$~mas, group delay residuals will generally be dominated by the effects 
of tropospheric (and at low frequencies ionospheric) mis-modeling, 
provided the geometric model for the array is accurate to better than $\pm1$ cm 
(including antenna locations, earth orientation parameters and solid-earth tides).
By observing $\approx10$ sources over a range of source azimuths and, most importantly, 
elevations in rapid succession ($\lax30$ min), one can estimate residual ``zenith'' 
tropospheric delays at each telescope.  These observing blocks are similar to those 
used for geodetic VLBI observations to determine source and telescope positions, 
as well as Earth's orientation parameters and, hence, are called ``geodetic blocks.''

The observing sequence can be determined 
by Monte Carlo simulations of large numbers of blocks and choosing the block
with the lowest expected zenith-delay uncertainties.   Operationally, 
it may be best to choose sources above an elevation of $\approx8^\circ$ and
available at a minimum of about $\sim60$\% of the telescopes in the array.  
Geodetic blocks should be placed before the start, during (roughly every 2 hours), 
and at the end of phase-reference observations.  This allows one to monitor slow 
changes in the total tropospheric delay for each telescope.
In order to measure group delays accurately, the observations should span the maximum
IF bandwidth, spaced in a ``minimum redundancy pattern'' to uniformly sample, as best 
as possible, all frequency differences.   For a system with 8 IF bands, this can be 
accomplished with bands spaced by 0, 1, 4, 9, 15, 22, 32 and 34 units.  If the 
recording system limited to $500$ MHz IF bandwidth, the unit separation would be 
$\approx14.7$ MHz.   With such a setup, estimated zenith delays 
are generally accurate to $\approx0.03$ nsec ($\approx1$~cm of propagation path-delay).
Comparisons of the geodetic block technique with those using Global 
Positioning System (GPS) data and an image-optimization approach
confirm this accuracy (\citealt{Honma:08b}).

Residual multi-band (group) delays and fringe rates are modeled as owing 
to a vertical tropospheric delay and delay-rate, as well as a clock
offset and clock drift rate, at each antenna.   
Note that the geodetic blocks need not be observed at the same frequency as the 
phase-referencing observations.  If one has independent calibration of ionospheric 
(dispersive) delays (see \S \ref{sect:ionosphere}), then it is simplest to observe the 
geodetic blocks at a frequency ($\nu$) above $\approx 20$ GHz to avoid contamination 
of the group delays by residual ionospheric effects, which decline approximately as 
$1/\nu^2$.  However, if one is limited to one (wide-band) receiver, then for observing 
frequencies below about $\approx10$ GHz, it is important to remove unmodeled 
ionospheric contributions to the geodetic group delays, since correction of 
interferometer phases has a different sign for dispersive compared to non-dispersive
delays.  This can be accomplished with ``dual-frequency'' geodetic blocks as outlined 
in \S\ref{sect:ionosphere}.

\subsubsection {GPS Tropospheric Delay Calibration:} \label{sect:GPS}

The GPS is a navigation system used to determine an
accurate three-dimensional position for an observer on the Earth.
The system consists of more than 20 satellites constituting
an artificial ``constellation'' of reference sources.
Accurate positions of the satellites are
broadcast by radio transmission to receivers on Earth.
The basic principles for position determination using GPS signals is similar 
to geodetic VLBI: the receiver position can be accurately determined based on 
delay measurements from multiple satellites.
Since the broadcasts are at low frequencies (1.2 to 1.6 GHz),
propagation delays in the ionosphere strongly affect the data.   
Dual frequencies allow removal of the ionospheric (dispersive) delay, and 
by observing several GPS satellites at different slant-paths through the atmosphere
simultaneously, one can solve for the tropospheric delay as well as the 
receiver position.  As such, one can use GPS data to calibrate tropospheric delays.
\citet{Steigenberger07} and \citet{Honma:08b} have conducted detailed comparisons of
tropospheric delay measurements by GPS and geodetic-mode VLBI (see \S \ref{sect:geoblocks})
observations; both studies conclude that the difference between the two methods
are small, $\lax2$ cm, with no systematic differences.

\subsection {Ionospheric Delay Calibration} \label{sect:ionosphere}

In principle, the effects of ionospheric (dispersive) delays can be largely 
removed by using global models of the total electron content (\citealt{Walker:00}),
by direct use of GPS data at each antenna, or from dual-frequency geodetic block 
observations.  Because the ionosphere is confined to a shell far above the Earth's
surface, rays from the source to a telescope can penetrate the ionosphere far from
the telescope location on the Earth's surface.  This generally favors using global models, 
based on smoothed results from a grid of GPS stations, over measurements made only at the
VLBI telescopes.   However, this is an area of current development and 
the methods of dispersive delay calibration may improve.

If one observes at frequencies below $\approx10$ GHz, one can in principle
obtain both dispersive and non-dispersive delays simultaneously from ``wide-band''
geodetic blocks.  For example, with receivers covering 4 to 8 GHz, one can
generate two separate sets of geodetic block data: 
spanning 500 MHz centered at the low end of the band ($\nu_L=4.25$ GHz) and at the
high end of the band ($\nu_H=7.75$ GHz).  
This gives observables  \tauLij\ and \tauHij, which contain both dispersive and
non-dispersive delays (eg, tropospheric delays and clock terms).
Differencing $\tauLij$ and $\tauHij$ values for all sources and baselines yields 
the dispersive contribution.   The dispersive contribution can be used to model the
ionospheric contribution and, importantly, can be scaled and subtracted from the 
$\tauHij$ delays to produce pure non-dispersive delays.   The non-dispersive delays
can then be modeled as described in \S\ref{sect:troposphere} to solve for pure
tropospheric and clock terms.  Note, that if one uses observations of a 
strong continuum source (fringe finder) to remove electronic delay differences
among IF sub-bands (\ie\ zeroing the delay on the fringe finder scan), 
this must be taken into account when modeling the dispersive delays.

An alternative to direct measurement is to use global maps of 
total ionospheric electron content generated from GPS data.
Global Ionosphere Maps (GIMs) are produced on a regular basis by several
groups, such as NASA, EAS, CODE, and UCP, who provide GIMs 
every two hours.  Basic trends in these global maps are generally consistent with
each other, although there are differences in scale for TEC values.
Also note that currently GIMs have a typical angular resolution of 2.5 deg, 
and thus small scale fluctuations in the ionosphere may not be well traced.
\citet{Hernandez09} compared TEC values obtained by GPS and direct measurements
(such as with JASON and TOPEX) and found vertical TEC values among the
methods differ at the $\approx20$\% level.
For a 50 TECU zenith column and an error of $\pm20$\%, the uncertainty in the 
ionospheric delay correction would be $\sim1$ cm at an observing
frequency of 22 GHz, which is slightly smaller than the delay error
caused by the troposphere.
On the other hand, at frequencies below 10 GHz the ionosphere 
is the dominant contributor to the delay error budget, even after 
corrections with GIMs are applied.

\subsection{Dual-beam Systems}

For a dual-beam receiving system, an additional calibration is 
required, because the propagation paths through the antenna and electronics 
are independent for the target and reference sources.
At VERA, the horn-on-dish method is utilized (\citealt{Honma:08a}), in which an
artificial noise source is mounted on an antenna's main reflector and a 
common signal is injected into both receivers.
During observations, cross-correlation of the noise source between the
two receivers is monitored in real-time, so that one can trace the time variation 
of the instrumental delay difference.  The measured delay difference between
the two receivers can be applied in post-processing.
\citet{Honma:08a} have conducted tests of the horn-on-dish calibration and
found that the VERA system can measure instrumental (path) delay differences 
to an accuracy of $\pm0.1$ mm, more than adequate for 10 \uas\ relative astrometry.

\subsection {Reference Source Position}

The error in the position of the source used as the phase reference causes 
an error in its delay/phase measurements, and these errors are then propagated 
to the target source.  The effect of a reference source position 
offset differs from other errors previously discussed.   The first order effect is that
the position offset of the reference source is transferred to the target source 
position.  If the reference source has negligible proper motion and parallax,
this only introduces a constant term, which is absorbed when fitting for
parallax and proper motion.  This is why we do not include such a term in 
Equation (\ref{eq:tau-obs}).

However, if the reference source position offset becomes large ($\gax10$ mas), 
there are additional second-order effects which come from the fact that the 
target and reference sources are separated on the sky. 
Because they are at different positions, the interferometer phase response to a 
position offset at one position on the sky does not exactly mimic the response at 
the other position.   These second order effects lead to a small position
error and to degraded image quality.
For an interferometric fringe spacing of 1 mas and a pair separation of $1^\circ$, 
one radian of second order phase-shift results from a calibrator position offset of 
$\sim10$ mas  (this offset, $\delta_\theta$, can be estimated by 
$2\pi \theta_{\rm sep} (\delta_\theta/\theta_{\rm beam}) \sim 1$).
In general, for phase-referencing astrometry, one needs to know the
calibrator position with an accuracy of $\lax10$ mas.
For a finer beam size or larger separation angle, the required positional
accuracy is even higher.

There are several methods for determining the absolute position of a phase reference 
source.  If one can find a compact ICRF radio source (with $\delta_\theta\approx1$ mas) 
close to the target source, one can transfer the position accuracy to the phase-reference
source.
However, ICRF sources are sparsely distributed on the sky and some are
heavily resolved on long interferometer baselines.  In such cases, one should include 
an ICRF source in the phase-referenced observations, even if offset by up to about 
$\approx5^\circ$ from the other sources used for differential astrometry, 
in order to determine their positions.  
Even if using a distant ICRF source as a reference produces poor images for other sources, 
one can usually obtain $\sim1$ mas accurate positions (and then discard the ICRF source data).

Alternatively, one can do preparatory observations to measure the position
of a phase reference source.   Usually connected-element interferometers (\eg\ JVLA,
eMERLIN, or the ATCA), which are limited in baseline length to $\lax100$ km,
can yield absolute positions with accuracies between 0.01 and 1 arcsec.
While this is usually adequate for VLBI data correlation, it is not 
good enough for high accuracy VLBI astrometry.  Instead, VLBI observations
using broad-band group-delay observables (as done for ICRF campaigns) are
preferable.  

If a maser source is to serve as the phase reference, then broad-band 
group-delays cannot be measured, since spectral lines are intrinsically very narrow,
leading to large uncertainty in group-delay estimates.
In this case, it is best to use an ICRF source to transfer positional accuracy
via phase-referencing.   As a last resort, if no qualified source is available, 
one can attempt to synthesize a maser spot image {\it without} phase-referencing.  
If the VLBI data have been correlated (or later corrected) with a model accurate 
to a few wavelengths of path delay, then such an image will resemble optical speckles 
(caused by short-term phase fluctuations induced by tropospheric water fluctuations 
and instabilities in the frequency standard).  However, the centroid of 
these speckles can often be determined to $\sim10$ mas, which may be sufficient 
for the position of the phase reference.

\subsection {Source Elevation Limits}

Often the dominant source of systematic error is uncompensated tropospheric
delays.   At any telescope, if one misestimates the zenith (\ie\ $z=0$) tropospheric 
delay by $\delta\tau_0$, this error is magnified at larger zenith angles by a
factor of $\approx\sec{z}$.   The differential effect between the target 
and a background source, with a zenith angle difference of $\Delta z$, is
then given by 
$\Delta\tau(z)=\delta\tau_0~\frac{\partial\sec{z}}{\partial z}\Delta z=\delta\tau_0\sec{z}\tan{z}\Delta z$.
Note that $\Delta\tau(z)$ increases dramatically at large zenith angles where both
$\sec{z}$ and $\tan{z}$ tend to diverge.  
For example, the ratio of differential 
delay error $\Delta\tau(z)$ at $z=70^\circ$ to $z=45^\circ$ is a factor of 8.0.
This suggests that astrometric observations should be limited, whenever possible, 
to small zenith angles.  Of course, when restricting observations to small zenith angles,
one must balance the loss of interferometer \uv-coverage, which affects both sensitivity and 
image fidelity, against increasing systematic errors that come with large zenith angle
observations.   In practice, it may prove valuable to fully simulate interferometric 
observations in order to estimate an optimum zenith angle limit for the declination of 
a given target source and the orientation and separation on the sky of the
background source.

\subsection {Measuring Positions}

Perhaps the simplest method of measuring a position from radio interferometric
data is to work in the image domain.   This involves placing the visibility 
\uv\ data on a grid, performing a 2-dimensional Fourier transformation
to sky coordinates, \uv\ $\rightarrow$ \xy, deconvolving the resulting
``dirty'' map using the ``dirty'' beam (point source response) with the
\clean\ algorithm, and fitting a 2-dimensional Gaussian 
brightness distribution to compact emission components.  This provides offsets, 
\dxdy, from the map center, whose absolute position is defined by the position used 
when cross-correlating the interferometer data and any subsequent shifts applied 
during calibration.

Care must be taken in both shifting positions and interpreting the measured
offsets from the map center, to properly account for the effects of 
precession, nutation, aberration and even general relativistic ray bending
near the Sun.  For example, position shifts applied to \uv-data are best done
by calculating the full interferometric delay and phase for the desired source position
and subtracting those values for the original (correlator) position.  Simply calculating
a differential correction can lead to significant astrometric errors. 
Similarly, one should not simply add map offsets, \dxdy\ (which ``know nothing''
of physical effects) to the map-center 
coordinates, without correcting for differential precession, nutation, and 
aberration, unless the map offsets are very small.  For example, over 
time scales of less than a few years, neglecting differential effects generally
leads to position errors of $\sim10^{-4}$\dxdy.  So, to ensure astrometric
accuracies of $\sim1$ \uas, \dxdy\ values should be $\lax10$ mas.
Therefore, if one measures large \dxdy\ values in a map, one should 
(accurately) shift positions in the \uv-data, re-image, and re-measure the
offsets.

\section {Future}

\subsection {Space VLBI}

Space-VLBI involves using a radio antenna in orbit around the
Earth to obtain baselines lengths greater than an Earth diameter.
As is clear from Equation \ref{eq:delay-error}, for fixed delay
uncertainty one gains astrometric accuracy as the baseline length increases.
For space VLBI to improve on Earth-based astrometry, the 
satellite position (orbit determination) must be determined to $\sim1$ cm 
accuracy, which is very challenging.   However, the complementary application
of using VLBI to accurately track spacecraft holds great promise (\citealt{Duev12}).

VSOP/HALCA was a space-VLBI mission launched in 1997.  Using VSOP,
\citet{Guirado01} conducted space-VLBI astrometry of the radio QSO pair
B1342+662/B1342+663 (which are only 5 arcmin apart) and demonstrated that the 
satellite position error was $\sim3$ meters and that the useful astrometry 
could be accomplished only for very close pairs.
Because of the poor performance of the VSOP 22 GHz receiving system,
maser astrometry was not attempted.

Currently the Russian space-VLBI satellite, RadioAstron, is in orbit 
(with a maximum interferometric baseline of $\sim 300,000$ km, 
comparable to the Earth-Moon separation).
Fringes from space-baselines have been obtained (\citealt{Kardashev:13}) and, 
perhaps, high-accuracy space-VLBI astrometry can be realized.

\subsection {The Event Horizon Telescope}

At millimeter and sub-mm wavelengths, VLBI can achieve an angular resolution 
sufficient to resolve event-horizon scales for nearby super-massive black holes.
A prediction of general relativity is that at this scale one should see
the ``shadow'' of the black hole (\citealt{Falcke:00}).  Impressive results
have been achieved using {\it ad hoc} arrays of antennas that can observe
at the short wavelengths required to ``see through'' a screen of electrons
that blurs the image of Sgr~A*, the super-massive black hole at the
center of the Milky Way (\citealt{Doeleman:08}).

The Event Horizon Telescope is a world-wide collaboration to realize
a powerful VLBI array operating at 1 mm or shorter wavelengths, anchored by
the phased-ALMA (acting as a single telescope with great collecting area).
In addition to imaging event horizon scales for Sgr~A* and M~87, one
should be able to explore the detailed structure of accretion disks and jet 
launching points in the vicinity of these super-massive black holes
using multi-frequency astrometry.
\citet{Broderick:11} investigated the possibility of such observations by
using the sub-array mode of ALMA, SMA, CARMA and other telescopes and demonstrated
that astrometry at the $\sim3$ \uas\ level is possible.  With such  
accuracy one can trace positional variations of the black holes owing to perturbations
from surrounding stars and/or a black hole companion.  In addition, one should be 
able to monitor structural variations occurring on dynamical timescales of the 
innermost stable circular orbit, which for Sgr~A* would be $\sim10$ min.

\subsection {Square Kilometer Array}

The Square Kilometer Array (SKA) is a world-wide project to build and
operate the next generation of radio interferometers with an
aggregate collecting area $\sim1$ km$^2$.  Achieving the
SKA will likely require three independent arrays, using different 
technologies to cover the frequency range of $\sim100$ MHz to $\sim20$ GHz.
Micro-arcsecond astrometry can be achieved at frequencies above $\sim3$ GHz,
provided baselines of several thousand km are an integral part of the design.
For astrometric observations, the SKA holds the promise of significantly
increased relative positional accuracy, because its great sensitivity allows
the use of weak calibrators much closer in angle on the sky to the 
astrometric target.  Compared to the VLBA, for example, the increased
sensitivity of the full SKA should allow the use of background compact 
radio sources at least an order of magnitude closer to the target,
resulting in astrometric accuracy better than $\pm1$~\uas!
In addition, if multi-beam feeds are used on individual antennas,
one could simultaneously observe many sources, greatly increasing astrometric 
survey speed.

\section*{Acknowledgment}

MH would like to thank Nobuyuki Sakai, Naoko Matsumoto,
Kazuhiro Hada, Hikaru Chida, Osamu Kameya, Fuyuhiko Kikuchi, and Yuka
Oizumi for their support in preparing the manuscript.
MH also acknowledges financial support by the MEXT/JSPS KAKENHI Grant Numbers 
24540242 and 25120007.

\end{document}